\documentclass[a4paper,11pt]{article}
\pdfoutput=1
\usepackage{jheppub}
\usepackage{amsmath,amssymb,array}
\usepackage{graphicx}
\usepackage{textcomp}
\usepackage[T1]{fontenc} 
\usepackage{color}
\usepackage{subfigure}
\usepackage{feynmp}
\usepackage{tabularray}
\usepackage{enumerate}
\usepackage{slashed}
\usepackage{hhline}
\usepackage{rotating}
\usepackage{hyperref}
\hypersetup{colorlinks=true,
	linkcolor=blue,
	filecolor=magenta,      
	urlcolor=blue}

\newcommand{\beqa}{\begin{eqnarray}}
\newcommand{\eeqa}{\end{eqnarray}}
\newcommand{\be}{\begin{equation}}
\newcommand{\ee}{\end{equation}}
\newcommand{\ba}{\begin{array}}
\newcommand{\ea}{\end{array}}

\title{\boldmath Loop-induced masses for the first two generations with optimum flavour violation}

\author[a,b]{Gurucharan Mohanta}
\author[a]{and Ketan M. Patel}

\affiliation[a]{Theoretical Physics Division, Physical Research Laboratory, \\Navarangpura, Ahmedabad-380009, India}
\affiliation[b]{Indian Institute of Technology Gandhinagar, Palaj-382055, India}

\emailAdd{gurucharan@prl.res.in}
\emailAdd{ketan.hep@gmail.com}

\abstract{A mechanism for the masses of third, second, and first generation charged fermions at the tree, 1-loop, and 2-loop levels, respectively, is proposed. The fermionic self-energy corrections that lead to this arrangement are induced through heavy vector bosons of a new gauged flavour symmetry group $G_F$. It is shown that a single Abelian group suffices as $G_F$. Moreover, the gauge charges are optimized to result in relatively smaller flavour violations in processes involving the first and second generation fermions. The scheme is explicitly implemented on the Standard Model fermions in an anomaly-free manner and is shown to be viable with observed charged fermion masses and quark mixings. Constraints from flavour violations dictate the lower limit on the new physics scale in these types of frameworks. Through optimal flavour violation, it is shown that nearly two orders of magnitude improvement can be achieved on the lower limit, leading to the new physics scale $\ge 10^3$ TeV in this case. Further improvements are possible at the cost of the down quark mass deviating more than $3 \sigma$ from its value extracted from lattice calculations. Options for inducing tiny masses for light neutrinos are also discussed.}

\begin{document}
\maketitle
\flushbottom

\section{Introduction}
\label{sec:intro}
The numerous and varied magnitudes of Yukawa couplings have been a perplexing and embarrassing feature of the Standard Model since its inception. Although technically natural, these couplings seem like an ad hoc arrangement to account for the observed flavours, their interchanging interactions, and their masses. It is desirable to have a mechanism that renders at least some of these parameters calculable. Radiative induction is one such possibility, in which some of these couplings are absent at the leading order in perturbation theory and are generated through loops \cite{Weinberg:1972ws,Georgi:1972hy}. This aligns with the observed mass pattern, as the mass hierarchies between two subsequent generations are roughly of the order of the loop suppression factor, $(4 \pi^2)^{-1}$.

What could be a minimal setup based on the radiative mass generation mechanism leading to a realistic flavour spectrum? Two requirements guide the efforts to answer this question. Firstly, the framework must lead to accidentally vanishing masses (or corresponding Yukawa couplings) for some generations at the leading order. Secondly, it must possess suitable interactions that induce their non-zero values through quantum corrections. It is straightforward to see that the Standard Model (SM) by itself is not sufficient to enable this. Even if the first $n$ generations ($n \leq 3$) are arranged to be massless by some means, the full action possesses a perturbative global $U(n)^5$ symmetry that prevents the generation of masses for these fermions at higher orders. This was realized in the very early attempts \cite{Georgi:1972hy,Mohapatra:1974wk,Barr:1978rv,Wilczek:1978xi,Yanagida:1979gs,Barbieri:1980tz} and highlighted the need for new interactions violating such a symmetry. Scenarios extending the SM based on new interactions, being mainly of Yukawa type \cite{Balakrishna:1987qd,Balakrishna:1988ks,Balakrishna:1988xg,Balakrishna:1988bn,Babu:1988fn,Babu:1989tv,Rattazzi:1990wu,Berezhiani:1991ds,Berezhiani:1992bx,Berezhiani:1992pj,Arkani-Hamed:1996kxn,Barr:2007ma,Graham:2009gr,Dobrescu:2008sz,Crivellin:2010ty,Crivellin:2011sj,Adhikari:2015woo,Chiang:2021pma,Chiang:2022axu,Baker:2020vkh,Baker:2021yli,Yin:2021yqy,Chang:2022pue,Zhang:2023zrn,Greljo:2023bix,Arbelaez:2024rbm,Greljo:2024zrj,Kuchimanchi:2024nkt} or gauge type \cite{HernandezGaleana:2004cm,Reig:2018ocz,Weinberg:2020zba,Jana:2021tlx,Mohanta:2022seo,Mohanta:2023soi} have been proposed. It has been observed that the latter typically provides a more concise and economical alternative from the perspective of overall calculability and predictivity. An extreme example of this is a framework proposed in \cite{Weinberg:2020zba}, in which the ambitious implementation turns out to be so predictive that it is discarded by the observed masses and mixing parameters of the charged fermions.

The realistic gauged extensions offering radiative masses for the first and second-generation fermions have been recently worked out in \cite{Mohanta:2022seo,Mohanta:2023soi}. The symmetry groups investigated as gauged flavour symmetries are abelian $G_F = U(1) \times U(1)$ \cite{Mohanta:2022seo} and non-abelian $G_F = SU(3)$ \cite{Mohanta:2023soi}. Two noteworthy outcomes common to these investigations are: (a) both the second and first generations receive non-zero masses at 1-loop itself, and their intergenerational mass hierarchy can be attributed to the gauge boson mass ordering; and (b) they are compelled to possess ${\cal O}(1)$ flavour violating couplings, forcing the new gauge bosons' masses $\sim\,10^8$ GeV or heavier. Both these features primarily depend on the structure of $G_F$. For example, the choices of $G_F$ in \cite{Mohanta:2022seo,Mohanta:2023soi} lead to more or less universal values of flavour changing neutral current couplings, namely $Q_{ij}$, between the $i^{\rm th}$ and $j^{\rm th}$ generations. On the other hand, the present experimental limits from quark and lepton flavour violations are more stringent in the $1$-$2$ sector than in the $2$-$3$ or $1$-$3$ sectors. This suggests that it would be desirable to have $|Q_{12}|<|Q_{23}|,|Q_{13}|$ to somewhat relax the lower limit mentioned in point (b) above. This has phenomenological and technical advantages in terms of observability and naturalness, respectively.

Given the points mentioned above, natural questions arise: Is there a choice of $G_F$ that induces the masses of only the second generation at one loop and the first generation at two loops? Can it accommodate flavour violation in an optimal amount such that the bound on the new gauge boson can be lowered further? The present work attempts to investigate these questions systematically. We begin with the simplest choice $G_F=U(1)$, as it extends the SM with only one gauge boson and obtains the conditions for the charges of the three generations of fermions under this symmetry, such that they obtain masses at successive orders in perturbation theory. In this way, an interesting correlation is found between the loop-induced first-generation mass and flavour violation in the $1$-$2$ sector, such that the latter vanishes completely in the limit of a massless first generation. By incorporating this mechanism for the SM fermions, the correlation is then exploited to achieve the optimal level of flavor violation while remaining consistent with the observed masses of the first-generation fermions in particular. The viable implementation requires a vectorlike pair of fermions for each charged fermion sector, along with multiple scalars carrying distinct charges under the $U(1)_F$ symmetry.

We begin with a toy framework in the next section and discuss conditions under which $G_F=U(1)_F$ can lead to masses for different generations at different orders in perturbation theory. Section \ref{sec:sym_dec} discusses an interplay between the loop-induced masses and flavour changing couplings at each order. In section \ref{sec:example_gauge}, we provide an explicit example of $U(1)_F$ charges demonstrating these features. The framework is then implemented in the standard model in a phenomenologically viable manner in section \ref{sec:SM} and constraints arising mainly from flavour violations are discussed in section \ref{sec:constraints}. We also discuss possibilities to accommodate massive neutrinos within this framework in section \ref{sec:neutrino} before concluding in section \ref{sec:summary}.

\section{The framework}
\label{sec:framework}
We first describe a mechanism leading to the third, second and first-generation fermion masses at tree, 1-loop and 2-loop levels, respectively, using a general framework. Consider three generations ($i=1,2,3$) of chiral fermions $f^\prime_{Li}$ and $f^\prime_{Ri}$ charged under a local flavour-dependent $U(1)_F$ symmetry with charges $q_{Li}$ and $q_{Ri}$, respectively. Additionally, consider a pair of vectorlike fermions, namely $F^\prime_{L,R}$, neutral under the $U(1)_F$. Under a chiral gauge symmetry like an electroweak symmetry of the SM, $f^\prime_{Li}$ and $f^\prime_{Ri}$ are to transform differently while $F^\prime_L$ and $F^\prime_R$  are to transform identically. Only the vectorlike fermion has a symmetry-preserving mass.

\subsection{At tree level}
Once the chiral and $U(1)_F$ symmetries are broken, the fermion mass Lagrangian is assumed to take the following form at the leading order:
\beqa \label{L_mass}
-{\cal L}_{m} &=& \mu_{L i}\, \overline{f}^\prime_{Li} F_R^\prime + \mu_{R i}\, \overline{F}^\prime_L f_{Ri}^\prime + m_F\, \overline{F}^\prime_L F^\prime_R + {\rm h.c.}\,, \nonumber \\
&\equiv & \overline{f}^\prime_{L \alpha}\, {\cal M}^{(0)}_{\alpha \beta}\, f^\prime_{R \beta} + {\rm h.c.}\,, \eeqa
with $f^\prime_{L(R)4} = F^\prime_{L(R)}$ and $\alpha = i,4$. The $4 \times 4$ matrix ${\cal M}^{(0)}$ is given by
\be \label{M0}
{\cal M}^{(0)} = \left(\ba{cc} 0_{3 \times 3} & (\mu_L)_{3 \times 1} \\ (\mu_R)_{1 \times 3} & m_F \ea \right)\,,\ee
with $\mu_L = (\mu_{L1},\mu_{L2},\mu_{L3})^T$ and $\mu_R = (\mu_{R1},\mu_{R2},\mu_{R3})$. The specific form of interactions in ${\cal L}_{m}$ can be arranged by promoting $\mu_{L,R}$ as spurions and assigning them appropriate charges under the full symmetry of the theory. Both $\mu_L$ and $\mu_{R}$ break $U(1)_F$, while at least one of them also breaks the chiral symmetry depending on the choice of the gauge charges of $F^\prime_{L,R}$. The structure of the tree-level mass matrix, eq. (\ref{M0}), closely resembles the form proposed in the context of the universal seesaw mechanism \cite{Berezhiani:1983hm,Berezhiani:1985in,Chang:1986bp}.

In the physical basis, obtained by $f_{L,R} = {\cal U}^{(0) \dagger}_{L,R} f^\prime_{L,R}$, one finds that only one of the chiral generation obtains non-vanishing mass due to the specific structure of ${\cal M}^{(0)}$. The bi-unitary diagonalisation leads to
\be \label{M0_diag}
{\cal U}_L^{(0) \dagger}\,{\cal M}^{(0)}\,{\cal U}_R^{(0)} \equiv {\cal D}^{(0)} = {\rm Diag}.\left( 0,0,m_3^{(0)},m_4^{(0)}\right)\,.\ee
An analytical simplification is possible in the so-called seesaw approximation, $\mu_{L,R} \ll m_F$. The unitary matrices can be expressed as
\be \label{U0_ss}
{\cal U}^{(0)}_{L,R} = \left(\ba{cc} U_{L,R}^{(0)} & - \rho_{L,R}^{(0)} \\
 \rho_{L,R}^{(0) \dagger} U_{L,R}^{(0)} & 1 \ea\right) + {\cal O}(\rho^2)\,, \ee
where
\be \label{rho0}
\rho^{(0)}_L = -\frac{1}{m_F} \mu_L\,,~~\rho_R^{(0) \dagger} = - \frac{1}{m_F} \mu_R\,,\ee
are three-dimensional column and row vectors, respectively. $U_{L,R}^{(0)}$ are $3 \times 3$ unitary matrices defined from
\be \label{M0_eff_diag}
U_L^{(0) \dagger}\, M^{(0)}\, U_R^{(0)} = {\rm Diag.}\left(0,0,m_3^{(0)}\right)\,.\ee
Here, $M^{(0)}$ is an effective $3 \times 3$ matrix obtained by integrating out the heavy vectorlike fermion pair. Explicitly,
\be \label{M0_eff}
M^{(0)}_{ij} = -\frac{1}{m_F}\, \mu_{L i}\, \mu_{R j}\,.\ee
As can be seen, $M^{(0)}$ is essentially a product of two vectors and, therefore, is a rank-1 matrix.

The simple form of $M^{(0)}$ allows one to determine the $U_{L,R}^{(0)}$ analytically. This can be obtained by finding the eigenvector corresponding to the only non-vanishing eigenvalue of $M^{(0)}$ and determining the other two eigenvectors from orthogonality. We find
\be \label{U0_ana}
U_L^{(0)} = \left( \ba{ccc} -\frac{\mu_{L2}^*}{\sqrt{N_1}} & -\frac{\mu_{L1} \mu_{L 3}^*}{\sqrt{N_2}} & \frac{\mu_{L1}}{\sqrt{N_3}} \\ 
\frac{\mu_{L1}^*}{\sqrt{N_1}} & -\frac{\mu_{L2} \mu_{L 3}^*}{\sqrt{N_2}} & \frac{\mu_{L2}}{\sqrt{N_3}} \\
0 & \frac{|\mu_{L1}|^2 + |\mu_{L2}|^2}{\sqrt{N_2}} & \frac{\mu_{L3}}{\sqrt{N_3}} \\
 \ea\right)\, V^{[12]}_L,\ee
where $N_{1,2,3}$ are normalization constants that can be determined by normalizing each of the three columns. $V^{[12]}_L$ represents an arbitrary unitary rotation in $1$-$2$ plane, which remain undetermined due to degeneracy among the first two masses. A similar expression for $U_R^{(0)}$ holds with replacement $\mu_L \to \mu_R^{*}$.

In the physical basis, the $U(1)_F$ gauge interactions are given by
\be \label{L_guage}
-{\cal L}_X = g_X X_\mu\, \left(({\cal Q}^{(0)}_L)_{\alpha \beta}\, \overline{f}_{L \alpha} \gamma^\mu f_{L \beta} + ({\cal Q}^{(0)}_R)_{\alpha \beta}\, \overline{f}_{R \alpha} \gamma^\mu f_{R \beta} \right)\,, \ee 
where
\be \label{Q0}
{\cal Q}_{L,R}^{(0)} = {\cal U}_{L,R}^{(0) \dagger}\,\left(\ba{cc} q_{L,R} & 0\\0 & 0 \ea \right)\,{\cal U}_{L,R}^{(0)}\,,\ee
and
\be \label{q_LR}
q_{L} = {\rm Diag.} \left(q_{L1},\,q_{L2},\,q_{L3}\right)\,,~~q_{R} = {\rm Diag.} \left(q_{R1},\,q_{R2},\,q_{R3}\right)\,.\ee
For flavour non-universal $q_{L,R}$, the gauge interactions are not flavour diagonal in the physical basis. This plays a pivotal role in inducing the masses of the remaining fermions at higher orders in perturbation theory.

\subsection{At 1-loop}
\begin{figure}[t]
\centering
\subfigure{\includegraphics[width=0.48\textwidth]{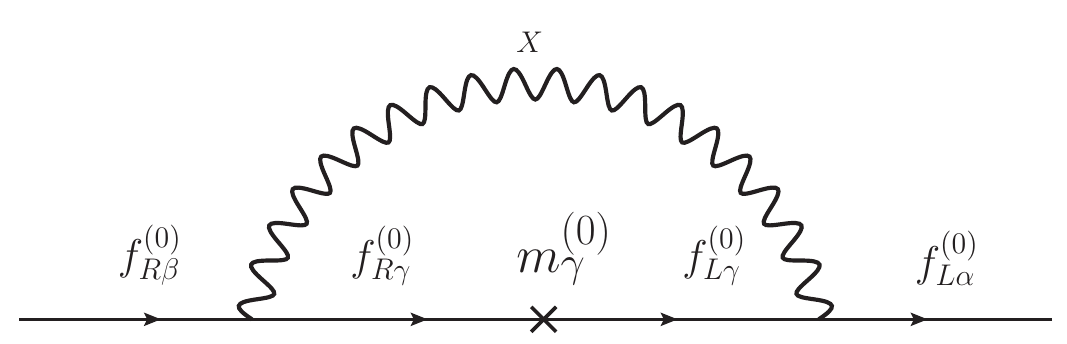}}\hspace*{0.5cm}
\subfigure{\includegraphics[width=0.48\textwidth]{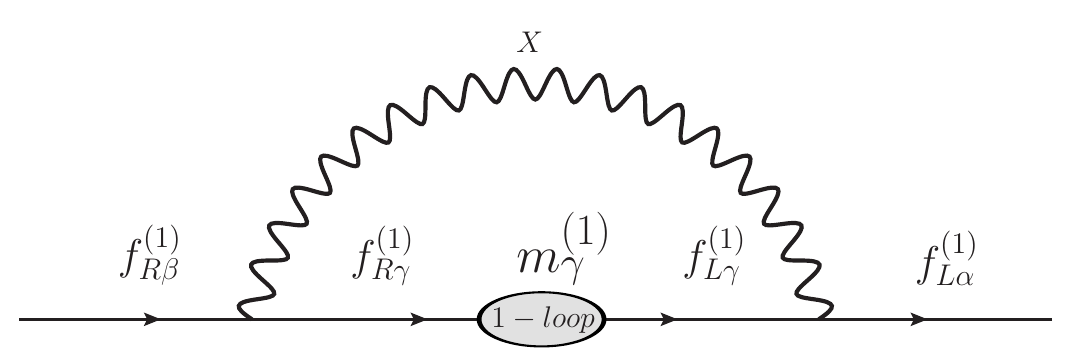}}
\caption{Loop diagrams that generate the masses of second-generation fermions (left panel) and first-generation fermions (right panel).}
\label{fig:loop}
\end{figure}

The 1-loop corrected fermion mass matrix can be parametrized as
\be \label{M1}
{\cal M}^{(1)} = {\cal M}^{(0)} + \delta {\cal M}^{(0)}\,, \ee
where
\be \label{dM0}
\delta {\cal M}^{(0)} = {\cal U}_L^{(0)}\,\sigma^{(0)}\,{\cal U}_R^{(0) \dagger}\,,\ee
and $\sigma^{(0)}$ captures contributions from the 1-loop diagrams involving fermions and the $X$-gauge boson in the loop, see the left panel in Fig. \ref{fig:loop}.
It is evaluated as (see for example \cite{Mohanta:2022seo})
\be \label{sigma_0}
\sigma^{(0)}_{\alpha \beta} = \frac{g_X^2}{4 \pi^2}\,\sum_\gamma ({\cal Q}^{(0)}_L)_{\alpha \gamma} ({\cal Q}^{(0)}_R)_{\gamma \beta}\, m_\gamma^{(0)}\, B_0[M_X,m_\gamma^{(0)}]\,.\ee
Here,
\be \label{B0}
B_0[m_a,m_b] = \Delta_\epsilon + 1 - \frac{m_a^2 \ln \frac{m_a^2}{Q^2} - m_b^2 \ln \frac{m_b^2}{Q^2}}{m_a^2 - m_b^2}\,,\ee
is a loop integration function evaluated in the dimensional regularization scheme at the renormalization scale $Q$, and 
\be \label{Delta_ep}
\Delta_\epsilon = \frac{2}{\epsilon} - \gamma_E + \ln 4\pi\,.\ee

Using eqs. (\ref{U0_ss},\ref{rho0},\ref{M0_eff_diag}) in (\ref{dM0}), it is straightforward to show that \cite{Mohanta:2022seo}
\be \label{dM0_eff}
\delta {\cal M}^{(0)}_{ij} \equiv \delta M^{(0)}_{ij} \simeq \frac{g_X^2}{4 \pi^2}\, q_{Li}\, q_{Rj}\, M^{(0)}_{ij}\, \left(B_0[M_X,m_3^{(0)}] - B_0[M_X,m_4^{(0)}] \right)\,,\ee
and $\delta {\cal M}^{(0)}_{\alpha 4} = \delta {\cal M}^{(0)}_{4 \alpha} = 0$. The vanishing fourth row and column of $\delta {\cal M}^{(0)}_{\alpha \beta}$ are due to the fact that the vectorlike fermions are chosen neutral under $U(1)_F$. The resulting ${\cal M}^{(1)}$, therefore, can be written as
\be \label{M1_1}
{\cal M}^{(1)} = \left( \ba{cc} \left(\delta M^{(0)}\right)_{3 \times 3} & \mu_L\\ \mu_R & m_F \ea \right)\,.\ee
The diagonalization of ${\cal M}^{(1)}$ leads to
\be \label{M1_diag}
{\cal U}_L^{(1) \dagger}\,{\cal M}^{(1)}\,{\cal U}_R^{(1)} \equiv {\cal D}^{(1)} = {\rm Diag}.\left( 0,m_2^{(1)},m_3^{(1)},m_4^{(1)}\right)\,.\ee
Here, $m_i^{(n)}$ is a mass of $i^{\rm th}$ fermion at $n^{\rm th}$ order. Interestingly, we find $m_1^{(1)}=0$ and only the second generation mass is induced at the 1-loop level. This is demonstrated below.

In the seesaw limit, the unitary matrices block diagonalizing ${\cal M}^{(1)} $ given in eq. (\ref{M1_1}) can be approximated as 
\be \label{U1_ss}
{\cal U}^{(1)}_{L,R} \approx \left(\ba{cc} U_{L,R}^{(1)} & - \rho_{L,R}^{(1)} \\
 \rho_{L,R}^{(1) \dagger} U_{L,R}^{(1)} & 1 \ea\right)\,, \ee
with $\rho^{(1)}_{L,R}=\rho^{(0)}_{L,R}$. Substituting the above in eq. (\ref{M1_diag}), one finds
the effective 1-loop corrected $3\times 3$ mass matrix
\be \label{M1_eff}
M^{(1)}_{ij} = M^{(0)}_{ij} + \delta M^{(0)}_{ij}\,,\ee
such that
\be \label{M1_eff_diag}
U_L^{(1) \dagger}\, M^{(1)}\, U_R^{(1)} = {\rm Diag.}\left(0,m_2^{(1)},m_3^{(1)}\right)\,.\ee
It can be noticed that the terms proportional to $\Delta_\epsilon$ in $\delta M^{(0)}_{ij}$ cancel, rendering the loop-induced mass finite and calculable.

We now show that $M^{(1)}$ always leads to one massless state. Substituting eq. (\ref{dM0_eff}) in (\ref{M1_eff}), it can be simplified to
\be \label{M1_eff_2}
M^{(1)}_{ij} = M^{(0)}_{ij}\,\left(1+ C\, q_{Li}\, q_{Rj} \right)\,, \ee
where $C= \frac{g_X^2}{4 \pi^2} (B_0[M_X,m_3^{(0)}] - B_0[M_X,m_4^{(0)}])$. It can then be seen that one of the columns of $M^{(1)}$ is not independent. For example,
\be \label{}
M^{(1)}_{i1} = \frac{q_{R1}-q_{R3}}{q_{R2}-q_{R3}}\, \frac{\mu_{R1}}{\mu_{R2}}\, M^{(1)}_{i2} + \frac{q_{R2}-q_{R1}}{q_{R2}-q_{R3}}\, \frac{\mu_{R1}}{\mu_{R3}}\, M^{(1)}_{i3}\,,\ee
for $i=1,2,3$. A similar relation also holds between the rows of $M^{(1)}$ with $q_R$ and $\mu_R$ replaced by $q_L$ and $\mu_L$, respectively. This implies, for the most general choices of $q_R$ and $q_L$, that the 1-loop corrected effective mass matrix is of rank-2 and, therefore, one of the states remains massless. In case of the degenerate $q_{L,R}$, $M^{(1)}_{ij}$ remains proportional to $M^{(0)}_{ij}$ leading to only one massive state as expected.

Although obtaining analytical expressions for $U_{L,R}^{(1)}$ can be a complicated exercise, the first columns of $U_{L,R}^{(1)}$ can be derived rather easily. We carry out this by finding the eigenvector corresponding to the vanishing eigenvalue of $M^{(1)} M^{(1) \dagger}$ and $M^{(1) \dagger} M^{(1)}$. The former gives the first column of $U_L^{(1)}$ while the latter gives the same of $U_R^{(1)}$.  We find,
\be \label{ev_m0}
\left(\ba{c} U^{(1)}_{L 11} \\ U^{(1)}_{L 21} \\ U^{(1)}_{L 31} \ea \right) = \frac{1}{\sqrt{N}} \left(\ba{c} 1 \\ \frac{\mu_{L1}^*}{\mu_{L2}^*} \frac{q_{L3} - q_{L1}}{q_{L2}-q_{L3}}\\ -\frac{\mu_{L1}^*}{\mu_{L3}^*} \frac{q_{L2} - q_{L1}}{q_{L2}-q_{L3}} \ea \right)\,.\ee
Similar expression with a replacement $L \to R$ holds for $U_{R}^{(1)}$. As can be seen, the massless state is a non-trivial combination of all three fermion flavours in general and, therefore, can pick up a mass at a higher order in perturbation theory in a similar way the second-generation mass is induced at 1-loop.

\subsection{At 2-loop}
Repeating the analysis carried out in the last subsection, we now estimate the first generation mass which gets induced through the next order correction to ${\cal M}^{(1)}$. It is captured by the diagram shown in the right panel of Fig. \ref{fig:loop}. Such a correction can be parametrized as
\be \label{M2}
{\cal M}^{(2)} = {\cal M}^{(1)} + \delta {\cal M}^{(1)}\,, \ee
with
\be \label{dM1}
\delta {\cal M}^{(1)} = {\cal U}_L^{(1)}\,\sigma^{(1)}\,{\cal U}_R^{(1) \dagger}\,,\ee
and $\sigma^{(1)}$ is given by an expression similar to eq. (\ref{sigma_0}). Explicitly,
\be \label{sigma_1}
\sigma^{(1)}_{\alpha \beta} = \frac{g_X^2}{4 \pi^2}\,\sum_\gamma ({\cal Q}^{(1)}_L)_{\alpha \gamma} ({\cal Q}^{(1)}_R)_{\gamma \beta}\, m_\gamma^{(1)}\, B_0[M_X,m_\gamma^{(1)}]\,,\ee
where
\be \label{Q1}
{\cal Q}_{L,R}^{(1)} = {\cal U}_{L,R}^{(1) \dagger}\,\left(\ba{cc} q_{L,R} & 0\\0 & 0 \ea \right)\,{\cal U}_{L,R}^{(1)}\,,\ee
is the charge matrix in the new physical basis obtained after 1-loop.

Again, one finds $\delta {\cal M}^{(1)}_{\alpha 4} = \delta {\cal M}^{(1)}_{4 \alpha} = 0$ for the same reason mentioned previously in the context of $\delta {\cal M}^{(0)}$. The correction to $3\times 3$ upper left block is explicitly computed as 
\beqa \label{dM1_33}
\delta{\cal M}^{(1)}_{ij} &=& \frac{g_X^2}{4 \pi^2}\, q_{Li}  q_{Rj}\, \left({\cal U}^{(1)}_L\right)_{i \gamma} \left({\cal U}^{(1)}_R\right)^*_{j \gamma}\,m_{\gamma}^{(1)}\,B_0[M_X,m_\gamma^{(1)}]\,.\eeqa
Altogether, ${\cal M}^{(2)}$ is given by
\be \label{M2_1}
{\cal M}^{(2)} = \left( \ba{cc} \left(\delta M^{(1)}\right)_{3 \times 3} & \mu_L\\ \mu_R & m_F \ea \right)\,,\ee
with 
\be \label{dM1_eff}
\delta M^{(1)}_{ij} = \delta M^{(0)}_{ij} + \delta{\cal M}^{(1)}_{ij}\,.\ee
Further simplification can be achieved in the seesaw approximation. Substituting eq. (\ref{U1_ss}) in eq. (\ref{dM1_33}) and after some straightforward algebraic simplification, we find
\beqa \label{dM1_33_1}
\delta{\cal M}^{(1)}_{ij} &=&  \frac{g_X^2}{4 \pi^2}\, q_{Li}  q_{Rj}\, \left(\sum_{k=2,3} (U_L^{(1)})_{ik} (U_R^{(1)})^*_{jk} m_k^{(1)} B_0[M_X,m_k^{(1)}] - M^{(0)}_{ij} B_0[M_X,m_F] \right)\,. \nonumber \\
\eeqa
This leads to
\beqa \label{dM1_eff_1}
\delta M^{(1)}_{ij} &=& \frac{g_X^2}{4 \pi^2}\, q_{Li} q_{Rj} \Big(M^{(0)}_{ij} (B_0[M_X,m_3^{(0)}]-2 B_0[M_X,m_F]) \Big. \nonumber \\
&+& \Big. \sum_{k=2,3}\,(U_L^{(1)})_{ik} (U_R^{(1)})^*_{jk} m_k^{(1)} B_0[M_X,m_k^{(1)}]\Big)\,,\eeqa
for the $3 \times 3$ matrix appearing in ${\cal M}^{(2)}$.

The diagonalization of ${\cal M}^{(2)}$ proceeds as
\be \label{M2_diag}
{\cal U}_L^{(2) \dagger}\,{\cal M}^{(2)}\,{\cal U}_R^{(2)} \equiv {\cal D}^{(2)} = {\rm Diag}.\left( m_1^{(2)},m_2^{(2)},m_3^{(2)},m_4^{(2)}\right)\,,\ee
leading to two-loop induced non-zero mass for the lightest fermion. As before the unitary matrices can be expressed as
\be \label{U2_ss}
{\cal U}^{(2)}_{L,R} \approx \left(\ba{cc} U_{L,R}^{(2)} & - \rho_{L,R}^{(2)} \\
 \rho_{L,R}^{(2) \dagger} U_{L,R}^{(2)} & 1 \ea\right)\,. \ee
The form of ${\cal M}^{(2)}$ in eq. (\ref{M2_1}) implies $\rho^{(1)}_{L,R}=\rho^{(0)}_{L,R}$. Using the above in eq. (\ref{M2_diag}), the effective $3\times 3$ mass matrix at 2-loop is obtained as
\be \label{M2_eff}
M^{(2)}_{ij} = M^{(0)}_{ij} + \delta M^{(1)}_{ij}\,,\ee
leading to
\be \label{M2_eff_diag}
U_L^{(2) \dagger}\, M^{(2)}\, U_R^{(2)} = {\rm Diag.}\left(m_1^{(2)},m_2^{(2)},m_3^{(2)}\right)\,.\ee

From eqs. (\ref{dM1_eff_1},\ref{M2_eff}) and some algebraic simplification using eq. (\ref{M1_eff_diag}), we finally obtain the following effective fermion mass matrix:
\beqa \label{M2_eff_fnl}
M^{(2)}_{ij} &=& M^{(0)}_{ij} \left(1+\frac{g_X^2}{4 \pi^2}\, q_{Li}\, q_{Rj}\, (B_0[M_X,m_3^{(1)}]- B_0[M_X,m_F]) \right) \nonumber \\
&+& \delta M^{(0)}_{ij} \left(1+\frac{g_X^2}{4 \pi^2}\, q_{Li}\, q_{Rj}\, B_0[M_X,m_3^{(1)}] \right) \nonumber \\
&+& \frac{g_X^2}{4 \pi^2}\, q_{Li}\, q_{Rj} \,(U_L^{(1)})_{i2} (U_R^{(1)})^*_{j2}\, m_2^{(1)}\, (B_0[M_X,m_2^{(1)}] - B_0[M_X,m_3^{(1)}])\,.\eeqa
The first term in the above is a usual tree-level contribution. The second and third terms parametrize the next-to-leading order contribution. These terms are proportional to $q_{L i} q_{R j} M^{(0)}_{ij}$. They, along with the tree-level contribution, lead to only two non-vanishing masses. The next-to-next-to-leading order effects arise from the fourth and fifth terms in eq. (\ref{M2_eff_fnl}) which gives rise to first-generation mass.

The singular part of $M^{(2)}$, as can be read from eq. (\ref{M2_eff_fnl}), is given by 
\beqa \label{div_M2}
{\rm Div.}\left(M^{(2)}_{ij}\right) & \propto & q_{Li}\, q_{Rj}\, \delta M^{(0)}_{ij} \propto q^2_{Li}\, q^2_{Rj}\, \left(U^{(0)}_{L}\right)_{i3}\,\left(U^{(0)}_{R}\right)^*_{j3}\,m_3^{(0)}\,. \eeqa
This contribution is of rank-1 and proportional to $m_3^{(0)}$. The divergent part can be removed through renormalizing $m_3^{(0)}$ as the latter is present as a parameter in the theory at tree level. The contribution to the first and second-generation masses arising from eq. (\ref{M2_eff_fnl}) is, therefore, finite and calculable.

\section{Symmetry deconstruction}
\label{sec:sym_dec}
It is insightful to understand how the assumed structure of the mass Lagrangian and gauge interactions lead to the masses for three generations of chiral fermions at successive orders in perturbation theory by inspecting the associated symmetries. As we show in this section, this helps in finding out the exact nature of the gauge interactions and the parameters in the mass Lagrangian which can lead to desired flavour spectrum.

First, in the absence of ${\cal L}_m$ and ${\cal L}_X$, the kinetic terms (along with any other flavour universal gauge interactions such as the ones present in the SM) are invariant under a global $U(3)_L \times U(3)_R$ symmetry. Nonvanishing $\mu_L$ and $\mu_R$ break this symmetry as it can be seen from ${\cal L}_m$ in eq. (\ref{L_mass}). However, one can choose an appropriate $U(3)_{L,R}$ rotations in order to bring $\mu_{L,R}$ in the form $(0,0,\times)$ where ``$\times$'' denotes some nonzero number. Therefore, the mass terms in eq. (\ref{L_mass}) leads to 
\be \label{tree_symm}
U(3)_L \times U(3)_R\, \xrightarrow{\mu_{L} \neq 0,\, \mu_{R} \neq 0}\, U(2)_L \times U(2)_R\,.\ee
It is this accidental $U(2)_L \times U(2)_R$ symmetry which gives rise to two massless states, identified as the first and second-generation fermions. In the physical basis, $f_{L,R} = {\cal U}^{(0) \dagger}_{L,R} f^\prime_{L,R}$, this symmetry of ${\cal L}_m$ can be seen more easily.

Next, if the full theory is invariant under this $U(2)_L \times U(2)_R$ then the nonzero masses for the lighter generations of fermions cannot be generated at any order in perturbation theory. In the present framework, breaking of this symmetry is achieved through the gauge interaction Lagrangian ${\cal L}_X$ given in eq. (\ref{L_guage}). The breaking\footnote{ Only for specific choices of $q_{L,R}$ leading to $Q_{L,R}^{(0)} = U_{L,R}^{(0) \dagger}\, q_{L,R}\, U^{(0)}_{L,R} = {\rm Diag.}\left(q,q,q^\prime\right) $, the gauge interactions in ${\cal L}_X$ do not break $U(2)_L \times U(2)_R$ symmetry. Here, $Q_{L,R}^{(0)}$ is upper-left $3\times3$ block of the matrix ${\cal Q}_{L,R}^{(0)}$ given in eq. (\ref{Q0}).} of $U(2)_L \times U(2)_R$ appears in the mass Lagrangian at $1$-loop. The corrected mass matrix $M^{(1)}_{ij}$, however, possesses again an accidental $U(1)_L \times U(1)_R$ subgroup of the original symmetry,
\be \label{1loop_symm}
U(2)_L \times U(2)_R\, \xrightarrow{\text{at 1-loop}}\, U(1)_L \times U(1)_R\,,\ee
leading to a massless state. Again, this symmetry is easier to express in the physical basis. It is defined by the transformations,
\be \label{U1U1_trans}
f_{L 1} \to e^{i \alpha_L} f_{L 1}\,,~~f_{R 1} \to e^{i \alpha_R} f_{R 1}\,.\ee

Subsequently, to give rise to non-vanishing mass for the first generation at higher loops involving $X$-boson, this symmetry must be broken in ${\cal L}_X$. In the new physical basis after 1-loop correction, the relevant charge matrix is:   
\be \label{QLR_1}
Q_{L,R}^{(1)} = U_{L,R}^{(1) \dagger}\, q_{L,R}\, U^{(1)}_{L,R}\,.\ee
Invariance of gauge interaction under transformations (\ref{U1U1_trans}) would imply $\left(Q_{L,R}^{(1)}\right)_{12} = \left(Q_{L,R}^{(1)}\right)_{13} = 0$. Non-vanishing $(Q_{L,R}^{(1)})_{12,13}$, therefore, are required to generate first-generation mass at 2-loop. Computing explicitly $Q_{L,R}^{(1)}$, we find
\beqa \label{Q_12}
\left(Q_L^{(1)}\right)_{12} &=& (q_{L3} - q_{L1})\, \left(U_L^{(1)}\right)^*_{31}\,\left(U_L^{(1)}\right)_{32} + (q_{L2} - q_{L1})\, \left(U_L^{(1)}\right)^*_{21}\,\left(U_L^{(1)}\right)_{22}\,,\nonumber \\
&=&\frac{(q_{L2}-q_{L1})(q_{L3}-q_{L1})}{\sqrt{N}\, (q_{L3}-q_{L2})}\,\left(\frac{\mu_{L1}}{\mu_{L3}} \left(U_L^{(1)}\right)_{32} -  \frac{\mu_{L1}}{\mu_{L2}} \left(U_L^{(1)}\right)_{22}\right)\,. \eeqa
Here, the first line is obtained using the orthogonality of the columns of $U_L^{(1)}$ while the second line follows from the expression eq. (\ref{ev_m0}). Similarly, 
\beqa \label{Q_13}
\left(Q_L^{(1)}\right)_{13} &=&\frac{(q_{L2}-q_{L1})(q_{L3}-q_{L1})}{\sqrt{N}\, (q_{L3}-q_{L2})}\,\left(\frac{\mu_{L1}}{\mu_{L3}} \left(U_L^{(1)}\right)_{33} -  \frac{\mu_{L1}}{\mu_{L2}} \left(U_L^{(1)}\right)_{23}\right)\,, \eeqa
and
\beqa \label{Q_23}
\left(Q_L^{(1)}\right)_{23} &=& (q_{L3}-q_{L2})\, \left(U_L^{(1)}\right)^*_{32}\left(U_L^{(1)}\right)_{33} - (q_{L2}-q_{L1})\, \left(U_L^{(1)}\right)^*_{12}\left(U_L^{(1)}\right)_{13}\,. \eeqa
Similar expressions can also be obtained for $\left(Q_R^{(1)}\right)_{ij}$ following the steps outlined above.

For $q_{L3} \neq q_{L2}$, it is noted that $(Q_L^{(1)})_{12,13}$ vanish in the limit: $q_{L1} \to q_{L2}$ or $q_{L1} \to q_{L3}$ or $\mu_{L1} \to 0$. Therefore, complete flavour non-degenerate $q_{L,R}$ and non-vanishing $(\mu_{L,R})_{1}$ are necessary to break the accidental $U(1)_L \times U(1)_R$ symmetry. The breaking of the latter appears at 2-loop and leads to suppressed but non-vanishing mass for the lightest fermion. Hence, at 2-loop,
\be \label{2loop_symm}
U(1)_L \times U(1)_R\, \xrightarrow{\text{at 2-loop}}\, U(1)_{\rm fn}\,,\ee
where $U(1)_{\rm fn}$ is a global fermion number which remains as an unbroken symmetry at least at the perturbative level in this setup.

Also, it can be noticed from the ${\cal L}_m$ that if any of the $\mu_{Li}$ or $\mu_{R i}$ is zero, then the corresponding $f^\prime_{Li}$ or $f^\prime_{Ri}$ does not mix with the rest of the fermions. As both the mass Lagrangian and gauge interactions are invariant under a $U(1)_L$ or $U(1)_R$ (under which $f^\prime_{L(R) i} \to e^{i \theta} f^\prime_{L(R) i}$), such a symmetry remains unbroken by the quantum corrections, leading to a massless fermion at all orders. Therefore, it is necessary to have all $\mu_{Li}$ and  $\mu_{Ri}$ nonvanishing in the present framework.

Two important observations that emerge from the above discussions are:
\begin{itemize}
\item Complete non-degeneracy in the flavour charges is necessary to induce the second generation mass at 1-loop and the first generation mass at 2-loop level. In addition, all $\mu_{Li}$ and  $\mu_{Ri}$ are required to be non-vanishing. As a consequence, the flavour changing current mediated through the $X$-boson is inherently present in the framework.
\item Up to 1-loop corrected mass Lagrangian, the strength of flavour changing current in $1$-$2$ and $2$-$3$ sectors, are parametrized by $\left(Q_L^{(1)}\right)_{12}$ and $\left(Q_L^{(1)}\right)_{23}$, respectively. Phenomenologically, it is desired to arrange $\left|\left(Q_L^{(1)}\right)_{12}\right| < \left|\left(Q_L^{(1)}\right)_{23}\right|$, as the experimental constraints from flavour violating processes in $1$-$2$ sector are relatively more severe. As can be seen from eq. (\ref{Q_12}), this can be arranged if 
\be \label{cond_1_qs}
|q_{L2}-q_{L1}| \ll |q_{L3}-q_{L2}|~~~{\rm or}~~~ |q_{L3}-q_{L1}| \ll |q_{L3}-q_{L2}|\,,\ee
and 
\be \label{cond_2_mus}
|\mu_{L 1}| < |\mu_{L2}|,\,|\mu_{L3}|\,.\ee 
The same applies to the parameters with $L \to R$ in the subscripts. 
\end{itemize}
These observations play a decisive role in fixing the exact nature of $U(1)_F$ as we discuss in the next section.

\section{An explicit example of gauge charges}
\label{sec:example_gauge}
We now discuss a possible choice for the flavour-dependent gauge charges and their implications on the loop-induced masses in the toy framework. Without losing much of the freedom in choosing flavour-specific charges, an easy way to ensure that the $U(1)_F$ is non-anomalous is to consider $U(1)_F$ as a vectorlike symmetry, i.e. $q_{L i} = q_{Ri}$ for each $i$. This choice is sufficient to cancel $[U(1)_F]^3$ and mixed gauge-gravity anomalies. Non-chiral nature of $U(1)_F$ by itself does not forbid the bare mass terms like $\overline{f}^\prime_{Li} f^\prime_{Ri}$ whose presence in ${\cal L}_m$ in eq. (\ref{L_mass}) can spoil the radiative mass generation mechanism and, therefore, an additional chiral symmetry may be needed. In the case of the SM, as we show in the next section, the gauge symmetry of the model itself can be utilised to serve this purpose.

Once $q_{Li} = q_{Ri}$ are chosen, one of the three charges can be fixed to a non-zero number without losing the generality. Based on this fact and the points summarized at the end of the previous section, we choose:
\be \label{gauge_charges}
q_{L1} = q_{R1} = 1 - \epsilon\,,~~~q_{L2} = q_{R2} = 1 + \epsilon\,,~~~q_{L3} = q_{R3} = -2\,, \ee
with $0< \epsilon \le 1$. Also, we have chosen the charges such that ${\rm Tr}(q_{L,R}) =0$. This makes it easier to cancel some of the mixed gauge anomalies when this toy framework is generalised to be integrated with the SM. For $\epsilon \ll 1$, the strength of flavour violation in $1$-$2$ sector can be relatively suppressed as argued in section \ref{sec:sym_dec}. Explicitly, for this choice, one finds from eqs. (\ref{Q_12},\ref{Q_23}):
\beqa \label{Q_rat}
\left| \frac{\left(Q_L^{(1)}\right)_{12}}{\left(Q_L^{(1)}\right)_{23}} \right| & \simeq & \frac{2 |\epsilon|}{3 \sqrt{N}}\, \left| \frac{\frac{\mu_{L1}}{\mu_{L3}} \left(U_L^{(1)}\right)_{32} -  \frac{\mu_{L1}}{\mu_{L2}} \left(U_L^{(1)}\right)_{22}}{ \left(U_L^{(1)}\right)^*_{32}\left(U_L^{(1)}\right)_{33}}\right|\,. \eeqa
When the condition, eq. (\ref{cond_2_mus}), is applied, $|\epsilon| \ll 1$ leads to $\left|\left(Q_L^{(1)}\right)_{12}\right| \ll \left|\left(Q_L^{(1)}\right)_{23}\right|$. However, for $\epsilon= 0$, one obtains a massless generation of fermions and therefore $\epsilon$ needs to be suitably optimized.

It can be seen that $\epsilon = 1$ reproduces an all-flavour version of the famous $L_\mu - L_\tau$ symmetry which we used for radiative generation of masses previously \cite{Mohanta:2022seo}.  Although, its generalisation in terms of charges given in eq. (\ref{gauge_charges}) looks ad-hoc, it can be shown to arise naturally in the effective theory from $U(1) \times U(1)$ gauge theory with well-defined charges and kinetic mixing. This is explicitly demonstrated in Appendix \ref{app:KM}. The free parameter $\epsilon$ can be associated with the kinetic mixing in this case. We also show that the underlying $U(1) \times U(1)$ symmetries, when combined with a flavour universal abelian symmetry like hypercharge, can be recast into $U(1)_1 \times U(1)_2 \times U(1)_3$ such that only one generation of fermions is non-trivially charged under each $U(1)$. Models of this variety have been recently proposed in \cite{FernandezNavarro:2023rhv,Davighi:2023evx} although not from the perspective of radiative mass mechanism.

\begin{figure}[t]
\centering
\subfigure{\includegraphics[width=0.48\textwidth]{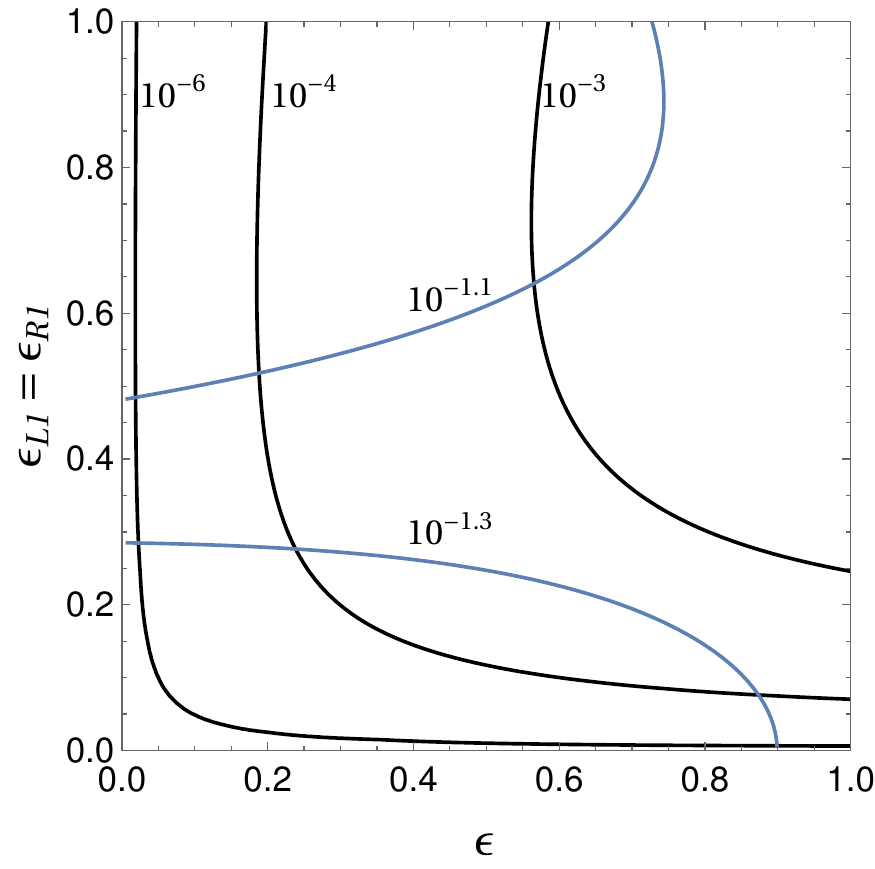}}\hspace*{0.5cm}
\subfigure{\includegraphics[width=0.48\textwidth]{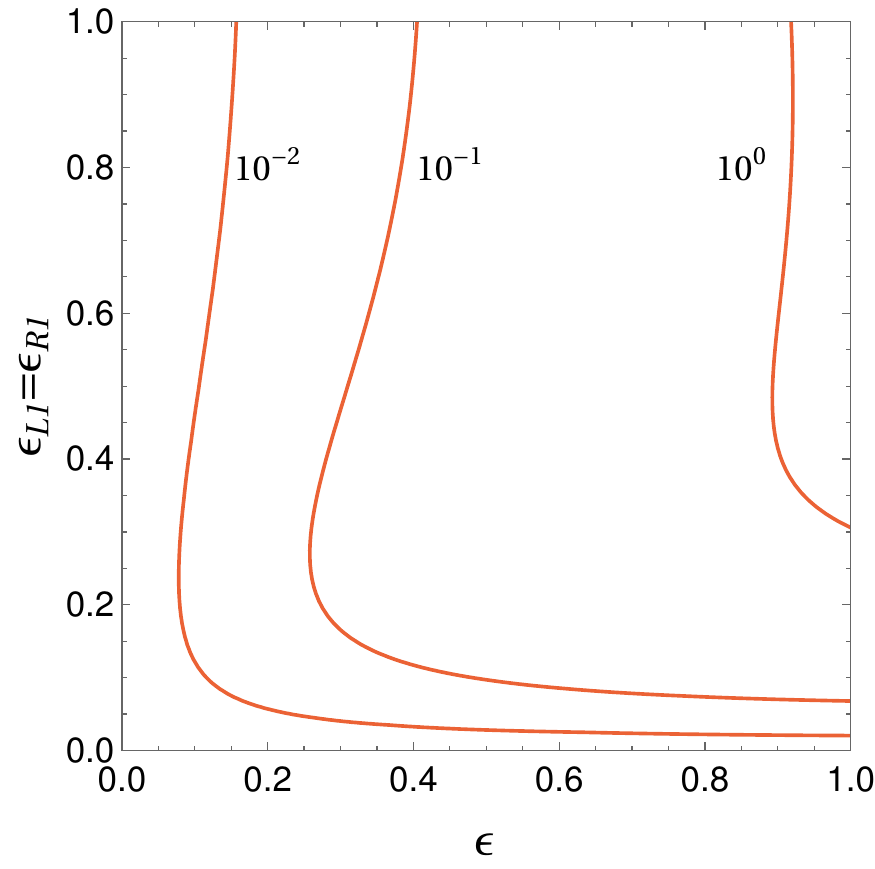}}
\caption{Left panel: contours of $m_1/m_3 = 10^{-6}$, $10^{-4}$, $10^{-3}$ (black) and $m_2/m_3 = 10^{-1.1}$, $10^{-1.3}$ (blue). Right panel: contours of $\left|Q^{(2)}_{12}\right|^2/\left|Q^{(2)}_{23}\right|^2 = 10^{-2}$, $0.1$ and $1$.}
\label{fig1}
\end{figure}
To explicitly demonstrate that the choice of gauge charges in eq. (\ref{gauge_charges}) leads to the desired pattern of fermion masses and flavour violating couplings, we numerically compute the 2-loop corrected mass matrix $M^{(2)}$ from eq. (\ref{M2_eff_fnl}) for an example set of input parameters. For the latter, we consider $g_X=0.5$, $M_X=10$ TeV, $m_F=10\, M_X$, $\mu_{L}= (\epsilon_{L1}, 0.3, 1)\, \frac{v}{\sqrt{2}}$, $\mu_{R}= (\epsilon_{R1}, 0.3, 1)$ TeV, $v=246$ GeV and $Q=M_Z$. The mass ratios $m_1/m_3$, $m_2/m_3$ for different values of $\epsilon$ and $\epsilon_{L 1} = \epsilon_{R 1} \equiv \epsilon_1$ are displayed in the left panel of Fig. \ref{fig1}. As it is anticipated, $m_1/m_3$ decreases as $\epsilon$ or $\epsilon_1$ become smaller. $m_2/m_3$ is not very sensitive to $\epsilon$ as the hierarchy between $m_2$ and $m_3$ is mostly governed by the loop factor. Similarly, the contours of the ratio $|Q^{(2)}_{12}|^2/|Q^{(2)}_{23}|^2$ are shown in the right panel of Fig.  \ref{fig1}.

\begin{figure}[t]
\centering
\subfigure{\includegraphics[width=0.46\textwidth]{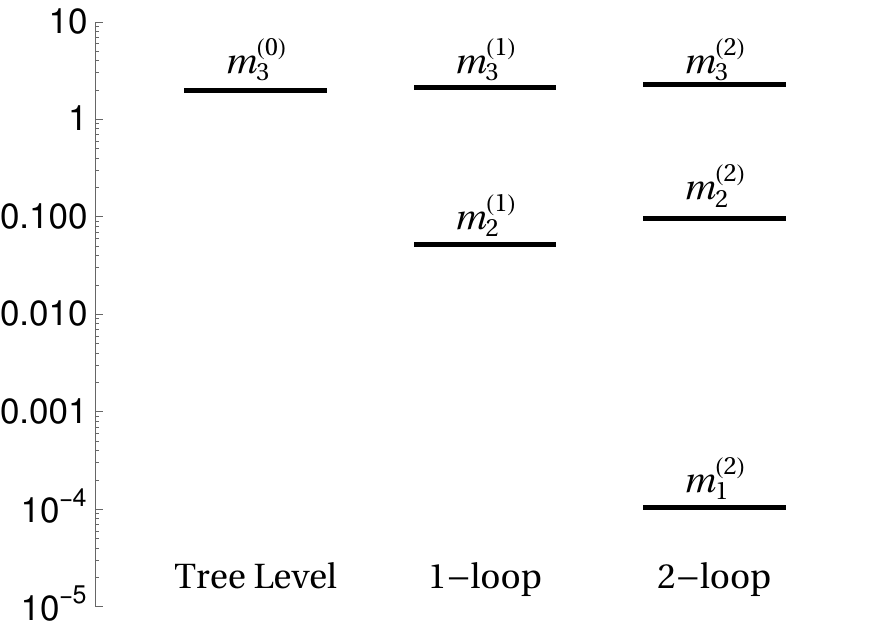}}\hspace*{0.5cm}
\subfigure{\includegraphics[width=0.46\textwidth]{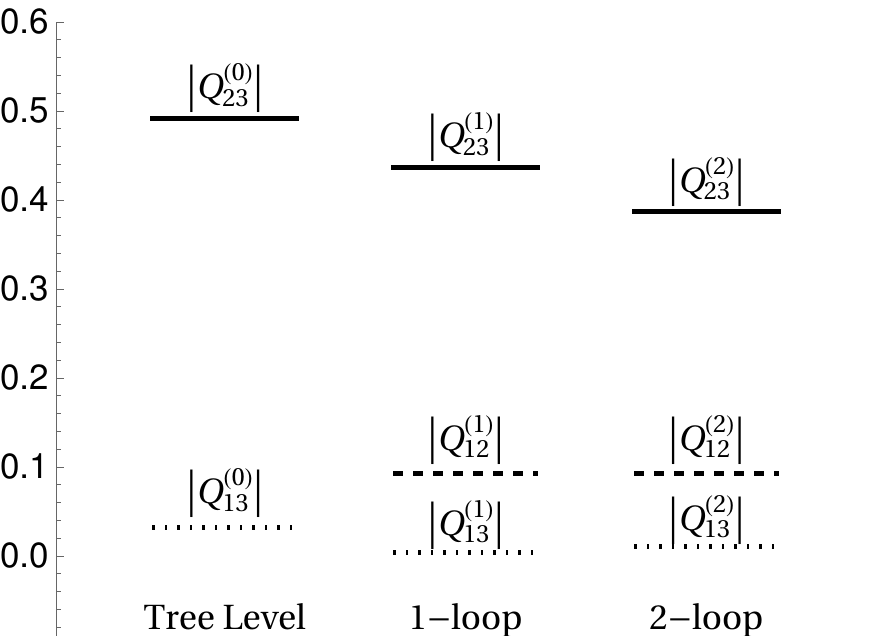}}
\caption{Left panel: Fermion masses at tree, 1-loop and 2-loop level for $\epsilon=\epsilon_{L1}=\epsilon_{R1} =0.2$ and for the remaining parameters as specified in the text. Right panel: For the same parameters, the magnitudes of $Q^{(n)}_{ij}$.}
\label{fig2}
\end{figure}
As it can be seen, $10^{-6} \gtrsim m_1/m_3 \gtrsim 10^{-4}$ typically prefers $\epsilon$ in the range $0.05$-$0.5$ for generic values of $\epsilon_1$. This leads to $|Q^{(2)}_{12}|^2 \lesssim 0.1\,|Q^{(2)}_{23}|^2$ as desired. Nevertheless, $|Q^{(2)}_{12}|$ cannot be made arbitrarily small as it requires very small $\epsilon$ and $\epsilon_1$. As discussed in the previous section and also confirmed by Fig. \ref{fig1}, this corresponds to the almost massless first generation. By fixing $\epsilon = \epsilon_1 = 0.2$, we also show the absolute values of three masses and flavour violating couplings $|Q^{(n)}_{ij}|$ in Fig. \ref{fig2} at $n^{\rm}$ loop. Due to degeneracy between $m_1$ and $m_2$ at the leading order, $|Q^{(0)}_{12}|$ is undefined. The latter gets fixed at 1-loop. Fig. \ref{fig2} shows that the masses and $|Q^{(n)}_{ij}|$ receive only small corrections at the next order and the ordering between them remains unchanged.

\section{Integration with the standard model}
\label{sec:SM}
This section discusses the integration of the above scheme with the SM and its viability in reproducing the observed charged fermion spectrum. We continue to work with the gauge charges as displayed in eq. (\ref{gauge_charges}) and apply them universally to all the SM fermions.

\subsection{Implementation}
\label{subsec:implementation}
It is straightforward to implement the underlying mechanism for the SM fermions, leading to loop-induced mass hierarchies for the charged fermions. The left chiral fields are generalised to include the electroweak quark and lepton doublets, $Q_{L i}$ and $L_{Li}$. Similarly, the right chiral SM fields include $u_{Ri}$, $d_{R i}$ and $e_{Ri}$. For the vectorlike fermions in each sector, we choose $U_{L,R}$, $D_{L,R}$ and $E_{L,R}$ which under the SM gauge symmetry transform in the same way as  $u_{Ri}$, $d_{R i}$ and $e_{Ri}$, respectively. The role of $\mu_{Li}$ is played by three pairs of electroweak doublets scalars $H_{u i}$ and $H_{d i}$  while terms proportional to $\mu_{R i}$ are to originate from three SM singlet scalars $\eta_{i}$.    The three generations of chiral fermions and scalars have flavour non-universal charges identical to the ones mentioned in eq. (\ref{gauge_charges}). The field content of the model and their charges under the SM and $U(1)_F$ gauge symmetries are summarized in Table \ref{tab:fields}. 
\begin{table}[!t]
\begin{center}
\begin{tabular}{ccc} 
\hline
\hline
~~Fields~~&~~$SU(3)_C \times SU(2)_L \times U(1)_Y$~~&~~$U(1)_F$~~~~\\
\hline
$Q_{L i} $ & $(3,2,\frac{1}{6}) $ & $(1-\epsilon, 1+\epsilon, -2)$ \\
$u_{R i} $ & $(3,1,\frac{2}{3}) $ & $(1-\epsilon, 1+\epsilon, -2)$ \\
$d_{R i} $ & $(3,1,-\frac{1}{3}) $ & $(1-\epsilon, 1+\epsilon, -2)$ \\
$L_{L i} $ & $(1,2,-\frac{1}{2}) $ & $(1-\epsilon, 1+\epsilon, -2)$ \\
$e_{R i} $ & $(1,1,-1) $ & $(1-\epsilon, 1+\epsilon, -2)$ \\
$\nu_{R i} $ & $(1,1,0) $ & $(1-\epsilon, 1+\epsilon, -2)$ \\
\hline
$U_{L,R}$ & $(3,1,\frac{2}{3}) $ & 0\\
$D_{L,R}$ & $(3,1,-\frac{1}{3}) $ & 0\\
$E_{L,R}$ & $(1,1,-1) $ & 0\\
\hline
$H_{u i}$ & $(1,2,-\frac{1}{2}) $ & $(1-\epsilon, 1+\epsilon, -2)$ \\
$H_{d i}$ & $(1,2,\frac{1}{2}) $ & $(1-\epsilon, 1+\epsilon, -2)$ \\
$\eta_i$  & $(1,1,0) $ & $(1-\epsilon, 1+\epsilon, -2)$ \\
\hline
\hline
\end{tabular}
\end{center}
\caption{List of fields and their flavour-universal SM and flavour-dependent $U(1)_F$ gauge charges.}
\label{tab:fields}
\end{table}

The presence of three generations of $\nu_R$ with $U(1)_F$ charges as listed in Table \ref{tab:fields} is essential to retain the vector structure of $U(1)_F$ symmetry which, as discussed in the previous section, ensures the cancellation of the cubic $U(1)_F$ and mixed gauge-gravity anomalies. It also implies vanishing $U(1)_Y \times U(1)_F^2$ anomaly. Moreover, the choice ${\rm Tr}(q_{L,R})=0$ leads to cancellation of anomalies associated with $SU(2)_L^2 \times U(1)_F$ and $U(1)_Y^2 \times U(1)_F$. The field content and gauge charges given in Table \ref{tab:fields}, therefore, lead to a theoretically consistent framework.

The renormalizable and gauge invariant Yukawa and mass Lagrangian involving the fermions can be written as
\beqa \label{LY_SM}
-{\cal L}_Y &=& {y_u}_i\,\overline{Q}_{Li}\, {H_u}_i\, U_R\, +\,{y_d}_i\,\overline{Q}_{Li}\, {H_d}_i\, D_R\,+\,  {y_e}_i\,\overline{L}_{Li}\, {H_d}_i\, E_R\,  \nonumber \\
& + & {y_u^{\prime}}_i\,\overline{U}_L\, \eta^*_i\, u_{R i}\, +\, {y_d^{\prime}}_i\,\overline{D}_L\, \eta^*_i\, d_{R i}\, +\, {y_e^{\prime}}_i\,\overline{E}_L\, \eta^*_i\, e_{R i}\, 
\nonumber \\
& + & m_U\, \overline{U}_L\,U_R\, + \,m_D\, \overline{D}_L\,D_R\, + \,m_E\, \overline{E}_L\,E_R\,+ \, {\rm h.c.}\,.\eeqa
As usual, the direct mass term for the three generations of fermions is forbidden by the chiral structure of the SM gauge symmetry even if they are allowed by $U(1)_F$. The non-trivial transformation assigned to the Higgs fields under the new gauge symmetry also forbids the direct Yukawa couplings between the left- and right-chiral fields of the SM.  The fermionic current associated with the flavour non-universal gauge interactions, ${\cal L}_X = j_X^\mu X_\mu$, is given by 
\be \label{JX_model}
j^\mu_X = g_X\, \left( \sum_{f = Q,L} q_{L i}\, \overline{f}_{Li}\,\gamma^\mu \, f_{L i}\, +\,  \sum_{f = u,d,e} q_{R i}\, \overline{f}_{R i}\,\gamma^\mu \, f_{R i} \right)\,,\ee
with the choice of $q_{L i}$ and $q_{R i}$ as listed in Table \ref{tab:fields}.

The most general non-vanishing vacuum of $\eta_i$ breaks the $U(1)_F$ completely. As usual, the electroweak symmetry is to be broken by the $U(1)_{\rm em}$ preserving vacuum expectation values (VEVs) of $H_{ui}$, $H_{di}$ which also contribute to the breaking of $U(1)_F$. We parametrize these VEVs as
\be \label{vev_def}
y_{fi}\,\langle H_{fi} \rangle \equiv \mu_{f i}\,,~~ y^\prime_{fi}\,\langle \eta^*_{i} \rangle \equiv \mu^\prime_{f i}\,,\ee
with $f = u,d,e$ and $\langle H_{ei}\rangle = \langle H_{di}\rangle$. The condition $\sum_i \left(|\langle H_{ui}\rangle|^2 + |\langle H_{di}\rangle|^2\right) = (246\,{\rm GeV})^2$ typically implies that $\mu_{fi}$ are at most of the order of the electroweak scale while there is no such restriction on the magnitudes of $\mu_{fi}^\prime$.

It can be seen that the substitution of eq. (\ref{vev_def}) in (\ref{LY_SM}) makes the latter take the precise form of the interactions depicted in eq. (\ref{L_mass}). The ${\cal M}^{(0)}$ given in eq. (\ref{M0}) has been replicated for $f  = u,d,e$ with $\mu_{Li} \to \mu_{fi}$ and $\mu_{Ri} \to \mu^\prime_{fi}$. Using the results derived in section \ref{sec:framework}, one finds the 2-loop corrected effective $3 \times 3$ matrix for the charged fermions as
\beqa \label{M2_eff_fnl_f}
\left(M^{(2)}_f\right)_{ij} &=& \left(M^{(0)}_f\right)_{ij} \left(1+\frac{g_X^2}{4 \pi^2}\, q_{Li}\, q_{Rj}\, (b_0[M_X,m_{f3}^{(1)}]- b_0[M_X,m_F]) \right) \nonumber \\
&+& \left(\delta M_f^{(0)}\right)_{ij} \left(1+\frac{g_X^2}{4 \pi^2}\, q_{Li}\, q_{Rj}\, b_0[M_X,m_{f3}^{(1)}] \right) \nonumber \\
&+& \frac{g_X^2}{4 \pi^2}\, q_{Li}\, q_{Rj} \,(U_{fL}^{(1)})_{i2} (U_{fR}^{(1)})^*_{j2}\, m_{f2}^{(1)}\, (b_0[M_X,m_{f2}^{(1)}] - b_0[M_X,m_{f3}^{(1)}])\,,\eeqa
for $f=u,d,e$. The corresponding mass matrices at the tree-level and one-loop are given by 
\be \label{M0_eff_f}
\left(M^{(0)}_f\right)_{ij} = -\frac{1}{m_F}\, \mu_{f i}\, \mu_{f j}^\prime\,,\ee
and 
\be \label{M1_eff_f}
\left(M^{(1)}_f\right)_{ij} = \left(M^{(0)}_f\right)_{ij}\,\left(1+ C_f\, q_{Li}\, q_{Rj} \right)\,, \ee
respectively, with 
$C_f= \frac{g_X^2}{4 \pi^2} (b_0[M_X,m_{f3}^{(0)}] - b_0[M_X,m_F])$.

Eqs. (\ref{M2_eff_fnl_f},\ref{M0_eff_f},\ref{M1_eff_f}) are straightforward generalisation of eqs. (\ref{M2_eff_fnl},\ref{M0_eff},\ref{M1_eff}). $m_{fi}^{(n)}$ appearing in these equations is the $i^{\rm th}$ eigenvalue of $M_f^{(n)}$ and can be computed using the given expressions. Similarly, $U^{(1)}_{fL}$ and $U^{(1)}_{fR}$ can be evaluated from the diagonalisation of $M_f^{(1)}$ following the usual definition, eq. (\ref{M1_eff_diag}). $b_0$ is the finite part of the function $B_0$ as defined in the $\overline{\rm MS}$ scheme and is given by
\be \label{b0}
b_0[m_a,m_b] = 1  - \frac{m_a^2 \ln \frac{m_a^2}{Q^2} - m_b^2 \ln \frac{m_b^2}{Q^2}}{m_a^2 - m_b^2}\,.\ee

 In the present model, the scalars can also give rise to radiative correction to the fermions mass matrices through diagrams similar to Fig. \ref{fig:loop} with $X$-boson replaced by the physical scalar fields. However, the form of interactions in eq. (\ref{LY_SM}) dictates that these contributions always depend on $H_{u,d}$-$\eta$ mixing. In turn, they depend on the several new parameters which are otherwise unconstrained and can spoil the calculability and predictivity offered by eq. (\ref{M2_eff_fnl_f}). Therefore, it would be desirable to suppress these contributions by assuming small or vanishing $H_{u,d}$-$\eta$ mixing. In Appendix \ref{app:scalar}, we outline conditions leading to this. In the subsequent analysis, we neglect the scalar-induced contributions and use eq. (\ref{M2_eff_fnl_f}) to quantify the quantum corrections to the fermion masses.

\subsection{Test of viability}
\label{subsec:numerical}
In order to demonstrate that $M^{(2)}_f$ obtained in eq. (\ref{M2_eff_fnl_f}) can lead to realistic charged fermion masses for $f=u,d,e$ and the quark mixing parameters, we carry out a numerical analysis following the method developed and described in detail in our previous work \cite{Mohanta:2022seo}. The essential points of the same are outlined in Appendix \ref{app:numerical} for the convenience of the reader.

We set $g_X = 0.5$ and optimize the $\chi^2$ function for several values of $\epsilon$ and a few example values of $M_X$. The remaining 23 parameters are left to be optimized considering some usual constraints, see discussion in Appendix \ref{app:numerical}. The entire exercise is carried out for the following two scenarios:
\begin{itemize}
\item Case A: Ordered $\mu_{fi}$ and $\mu_{fi}^\prime$, i.e. for $f=u,d,e$, we impose
\be \label{caseA}
|\mu_{f1}| < |\mu_{f2}| < |\mu_{f3}|\,,~~{\rm and}~~~|\mu_{f1}^\prime| < |\mu_{f2}^\prime| < |\mu_{f3}^\prime|\,.\ee
\item Case B: Strongly ordered $\mu_{di}$ and $\mu_{di}^\prime$. In this case in addition to the conditions in eq. (\ref{caseA}), we impose
\be \label{caseB}
\frac{|\mu_{d1}|}{ |\mu_{d2}|} < 0.1\,,~~{\rm and}~~~\frac{|\mu_{d1}^\prime|}{ |\mu_{d2}^\prime|}<0.1\,.\ee
\end{itemize}
Both these conditions help in achieving phenomenologically favoured ordering in flavour violations as discussed in section \ref{sec:example_gauge} and provide a transparent connection between the flavour violation in $1$-$2$ sector and the parameter $\epsilon$. Case B, in particular, points out a parameter space in which the strongest constraints from the $K^0$-$\overline{K}^0$ oscillations can be evaded more effectively, as we show later in the next section.

\begin{figure}[t] 
\centering
\subfigure{\includegraphics[width=0.48\textwidth]{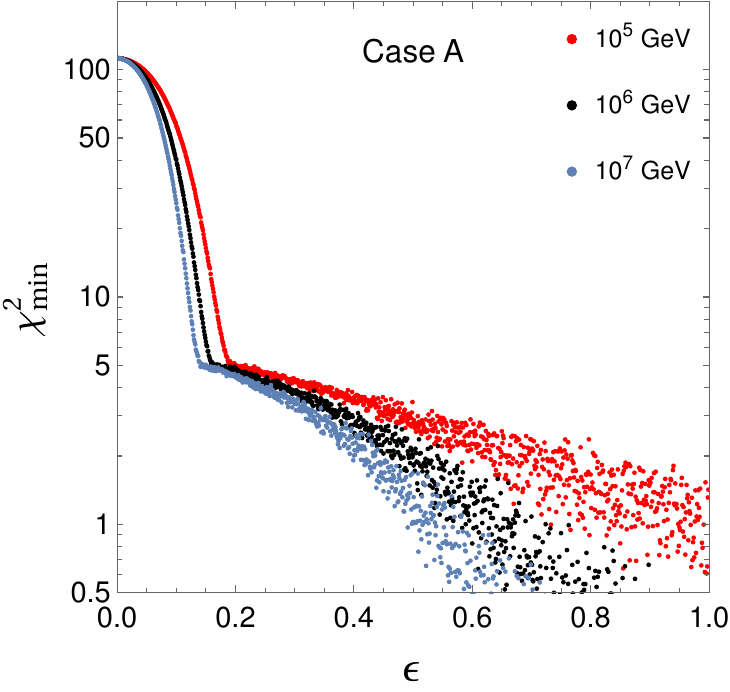}}\hspace*{0.5cm}
\subfigure{\includegraphics[width=0.48\textwidth]{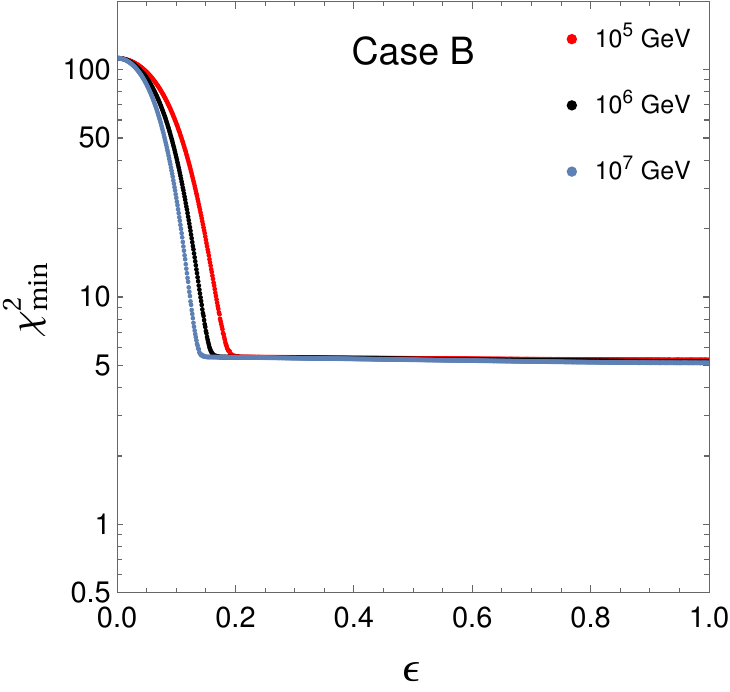}}
\caption{The obtained $\chi^2_{\rm min}$ as a function of $\epsilon$ for three sample values of $M_X$ with conditions categorized under case A (left panel) and case B (right panel). }
\label{fig:chi2}
\end{figure}

The results of minimized $\chi^2$ for different values of $\epsilon$ and $M_X$ for both the above cases are displayed in Fig. \ref{fig:chi2}. 
The fits corresponding to $\chi^2_{\rm min} \le 9$ can be considered acceptable, as none of the observables is more than $3 \sigma$ away from its mean value. Keeping this in consideration, the noteworthy observations from Fig. \ref{fig:chi2} are the following.
\begin{itemize}
\item The fits in all cases disfavour $\epsilon < 0.15$. We find that for the smaller values of $\epsilon$, the first-generation masses do not get fitted to their experimentally extracted values within their $3 \sigma$ ranges. This is in agreement with expectation discussed in section \ref{sec:example_gauge}. 
\item In case A, the fit improves with larger values of $\epsilon$. In fact, for $M_X \ge 10^6$ GeV, excellent fits can be achieved corresponding to $\chi^2_{\rm min} \lesssim 1$ for $\epsilon > 0.5$ implying that all the observables can be fitted within their $1\sigma$ ranges. Also, improvement in fits for a given $\epsilon$ is observed with larger $M_X$. 
\item  In contras to the above, $\chi^2_{\rm min} < 5$ cannot be achieved for any value of $\epsilon$ and $M_X$ in case B. We find that with conditions in eq. (\ref{caseB}) the mass of the down quark remains $2.3 \sigma$ away from its central value irrespective of the other parameters. 
\end{itemize}
From the above observations, it is evident that the mechanism can be implemented on the SM in a phenomenologically consistent manner only if $\epsilon > 0.15$. This exact lower limit then implies a finite amount of flavour violation in the $1$-$2$ sector, which we discuss in detail in the next section.

To demonstrate the efficiency of the fits and behaviour of the fitted parameters,  we give two explicit benchmark solutions in Table \ref{tab:fit}. 
\begin{table}[t]
\begin{center}
\begin{math}
\begin{tabular}{cccccccc}
\hline
\hline
 &   &
      \multicolumn{2}{c}{\bf Solution 1} &
      \multicolumn{2}{c}{\bf Solution 2} \\
~~~Observable~~~ & $O_{\rm exp}$  & $O_{\rm th}$ & Pull & $O_{\rm th}$ & Pull  \\
\hline
$m_u$\, [MeV]  &  $1.27 \pm 0.5$   &$ 1.26$    & $ -0.02$ & $ 1.27$     &  $ 0$ \\
$m_c$\, [GeV]   & $0.619 \pm 0.084$    &$ 0.614$    & $ -0.06$ & $ 0.617$     &  $-0.02$ \\
$m_t$\, [GeV]   & $171.7 \pm 3.0$     & $ 171.8$  & $ 0.03$ & $ 171.7 $   & $ 0$  \\
$m_d$\, [MeV]  & $2.90 \pm 1.24$& $ 0.16$  & $ -2.21$  & $0.02$ &$ -2.32$\\
$m_s$\, [GeV]  & $0.055 \pm 0.016$      & $ 0.056$  & $0.06 $ & $ 0.055$    & $0$ \\
$m_b$\, [GeV]  & $2.89 \pm 0.09$         & $ 2.89$     & $0$ & $ 2.89$       & $0 $ \\
$m_e$\, [MeV]  & $0.487 \pm 0.049$        & $ 0.489$   & $0.04$ & $ 0.487$     & $0$  \\
$m_{\mu}$\,[GeV] &$0.1027 \pm 0.0103$      & $ 0.1025$  & $-0.02$  & $ 0.1025$     & $-0.02 $ \\
$m_{\tau}$ \,[GeV] & $1.746 \pm 0.174$       & $ 1.746$   & $0$ & $ 1.742$      & $-0.02$ \\
$|V_{us}|$     & $0.22500 \pm 0.00067$      & $0.22499$    & $-0.01 $ & $0.22500$     & $0$  \\
$|V_{cb}|$     & $0.04182 \pm 0.00085$          & $0.04182$   & $0$& $0.04182$    & $0$  \\
$|V_{ub}|$     & $0.00369 \pm 0.00011$          & $0.00369 $   & $0 $ & $0.00369$    & $0$  \\
$J_{\rm CP}$   & $3.08\times10^{-5}$   & $3.08\times10^{-5}$  & $0 $ & $3.08\times10^{-5}$ & $0$ \\
\hline
    $\chi^2_{\rm min}$    & & & $4.9$ & & $5.4$ \\
\hline
\hline
\end{tabular}
\end{math}
\end{center}
\caption{Two benchmark solutions of best-fit from cases A and B, respectively. $O_{\rm exp}$ denote the extrapolated values of the respective observables at the renormalization scale $Q=M_Z$ . The reproduced values of the observables at $\chi^2_{\rm min}$ are listed under $O_{\rm th}$ and the corresponding pulls are given which indicates the amount of deviation from the mean  $O_{\rm exp}$ value.}
\label{tab:fit}
\end{table}
The Solution 1 (S1) falls within case A. Solution 2 (S2) corresponds to strongly ordered $\mu_{di}$ and $\mu_{di}^\prime$ in case B. They are also chosen with optimized values of $\epsilon$ and $M_X$ from the perspective of flavour violation, as will be revealed in the next section. It can be seen that all the observables, except $m_d$, are fitted excellently. $m_d$ remains more than $2 \sigma$ away in both the cases. Therefore, future improvements in the uncertainty in $m_d$ can have a significant effect on the viability of these types of solutions.

The values of $\epsilon$ and $M_X$ for the chosen benchmark solutions and numerical values of all the remaining 25 real parameters at the minimum of $\chi^2$ are listed in Table \ref{tab:sol} in Appendix \ref{app:numerical}. The most noteworthy aspect of the values of the various parameters in both cases is that all $\mu_{fi}$ and $\mu_{fi}^\prime$ are spread in the range of only two orders of magnitude at most. This, when interpreted in terms of eq. (\ref{vev_def}), implies that all the fundamental Yukawa couplings can be of ${\cal O}(1)$ in this kind of model. Obviously, the hierarchies in the observed masses are then attributed to the very intricate and carefully arranged structure of the theory. It is also noticed that the vectorlike fermions must stay close to the $U(1)_F$ breaking scale. This is to avoid large seesaw suppression in the masses of the third-generation fermions. This is more apparent in the case of top quark mass, which forces $m_U \simeq \mu_{u3}^\prime$. Relatively light $m_b$ and $m_\tau$ are arranged through $m_D > \mu_{d3}^\prime$ and $m_E > \mu_{e3}^\prime$, respectively.

\section{Phenomenological constraints}
\label{sec:constraints}
The most significant constraint on the new physics in the present setup is expected to come from flavour changing neutral current (FCNC) interactions of the SM fermions. They can arise in three ways: (i) through a direct mediation by $X$-boson which has flavour changing couplings, (ii) through $Z$-boson and mixing of the SM fermions with the vectorlike ones, and (iii) through a mediation by neutral scalars. The first is parametrized by eq. (\ref{JX_model}) in the physical basis of the quarks and charged leptons as
\be \label{JX_model_phys}
j^\mu_X = g_X\, \sum_{f = u,d,e} \left(\left(X^f_L\right)_{ij}\, \overline{f}_{Li}\,\gamma^\mu \, f_{L j}\, +\,  \left(X^f_R\right)_{ij}\, \overline{f}_{R i}\,\gamma^\mu \, f_{R j} \right)\,,\ee
with 
\be \label{Xf}
X^f_{L,R} = U^{f \dagger}_{L,R}\, q_{L,R}\, U^f_{L,R}\,,\ee
and $U^{f}_{L,R}$ being the $3 \times 3$ unitary matrices that diagonalize the corresponding 2-loop corrected $M^{(2)}_f$ given in eq. (\ref{M2_eff_fnl_f}) such that $U^{f \dagger}_{L}\, M^{(2)}_f\, U^f_{R} = {\rm Diag.}(m^{(2)}_{f1},m^{(2)}_{f2},m^{(2)}_{f3})$.

The current associated with the $Z$-boson in the physical basis is obtained as
\be \label{JZ_model_phys}
j^\mu_Z = -\frac{g}{2 \cos \theta_W}\, \left({\cal Y}^u_{\alpha \beta}\, \overline{u}_{L \alpha}\gamma^\mu u_{L \beta}\, -\, {\cal Y}^d_{\alpha \beta}\, \overline{d}_{L \alpha}\gamma^\mu d_{L \beta}\, -\, {\cal Y}^e_{\alpha \beta}\, \overline{e}_{L \alpha}\gamma^\mu e_{L \beta}\, -2 \sin^2 \theta_W\, J_{\rm em}^\mu \right)\,,\ee
where $J_{\rm em}^\mu$ is flavour diagonal electromagnetic current. The $4 \times 4$ coupling matrices are given by
\be \label{Yf}
{\cal Y}^f = {\cal U}_L^{f(2) \dagger}\, \left(\ba{cc} {\bf 1}_{3 \times 3} & 0 \\ 0 & 0 \ea \right)\,{\cal U}_L^{f(2)}\,. \ee 
 Using the seesaw expansion, eq. (\ref{U2_ss}), one finds
\be \label{Yf_eff}
{\cal Y}^f_{ij} \simeq \delta_{ij} + {\cal O}(\frac{\mu_f\, \mu^\prime_f}{m_F^2})\,. \ee
 Apparently, the off-diagonal elements are suppressed by heavy-light mixing. For the values of $\mu_f$ and $m_F$ given Table \ref{tab:sol}, one finds that the FCNCs induced by the $Z$-boson remain suppressed in comparison to that induced by the $X$-boson. 
 
Similarly, the flavour changing couplings of the neutral scalars are also found suppressed by the heavy-light mixing. This is demonstrated in detail in Appendix \ref{app:scalar}. Consequently, if the scalar masses are of ${\cal O}(M_X)$, the FCNC's induced by the scalars provide only the sub-leading contributions to the ones mediated by the $U(1)_X$ gauge boson. Therefore, we focus on the current parametrized by eq. (\ref{JX_model_phys}) and derive the constraints on $M_X$ arising from the following most relevant observables.

\subsection{Meson-antimeson oscillation}
The most stringent limits on quark sector FCNC couplings arise from the neutral meson-antimeson oscillations, namely $M^0$-$\overline{M}^0$ transitions, where $M=K,\, B_d,\, B_s,\, D$. To evaluate these constraints in the context of the present model, we follow the methodology developed and discussed in detail in our previous work \cite{Mohanta:2022seo}. The effective four-fermion operators obtained by eliminating the $X$-boson are given by
\be \label{FV_op}
{\cal L}^{\rm eff}_X = -\frac{1}{M_X^2}\,j^\mu_X\,j_{X \mu}\,,\ee 
 where $j^\mu_X$ is already specified in eq. (\ref{JX_model_phys}). This can be compared with the effective Hamiltonian, ${\cal H}^{\rm eff}_M\,=\, \sum_{i=1}^5 C_M^i Q^i + \sum_{i=1}^3 \tilde{C}_M^i \tilde{Q}_i$ that parametrizes $\Delta F = 2$, $M^0$-$\overline{M}^0$ transitions in terms of the familiar Wilson coefficients \cite{UTfit:2007eik}. Through this, we can identify the respective Wilson coefficients $C^i_M$ and $\tilde{C}^i_M$ in terms of flavour violating couplings present in the underlying model. At the matching scale $Q=M_X$, we find:
\beqa \label{WCs}
C_K^1 &=&{\frac{g_X^2}{M_X^2}\left[\left(X^d_L\right)_{12}\right]^2}\,,~~\tilde{C}_K^1=\frac{g_X^2}{M_X^2}\left[\left(X^{d}_{R}\right)_{12}\right]^2\,,~~C_K^5=-4\frac{g_X^2}{M_X^2}\left(X^{d}_{L}\right)_{12} \left(X^{d}_{R}\right)_{12}\,. \nonumber\\
C_{B_d}^1 &=& \frac{g_X^2}{M_X^2}\left[\left(X^{d}_{L}\right)_{13}\right]^2\,,~~\tilde{C}_{B_d}^1=\frac{g_X^2}{M_X^2}\left[\left(X^{d}_{R}\right)_{13}\right]^2\,,~~C_{B_d}^5=- 4\frac{g_X^2}{M_X^2}\left(X^{d}_{L}\right)_{13} \left(X^{d}_{R}\right)_{13}\,,\nonumber\\
C_{B_s}^1 &=& \frac{g_X^2}{M_X^2}\left[\left(X^{d}_{L}\right)_{23}\right]^2\,,~~\tilde{C}_{B_s}^1=\frac{g_X^2}{M_X^2}\left[\left(X^{d}_{R}\right)_{23}\right]^2\,,~~C_{B_s}^5=- 4\frac{g_X^2}{M_X^2}\left(X^{d}_{L}\right)_{23} \left(X^{d}_{R}\right)_{23}\,,\nonumber\\
C_D^1 &=& \frac{g_X^2}{M_X^2}\left[\left(X^{u}_{L}\right)_{12}\right]^2\,,~~\tilde{C}_D^1=\frac{g_X^2}{M_X^2}\left[\left(X^{u}_{R}\right)_{12}\right]^2\,,~~C_D^5=- 4\frac{g_X^2}{M_X^2}\left(X^{u}_{L}\right)_{12} \left(X^{u}_{R}\right)_{12}\,.\eeqa
The rest of the $C^i_M$ and $\tilde{C}^i_M$ are vanishing at $Q=M_X$.

Next, all the coefficients are evolved using appropriate renormalization group equations (RGE) from $Q=M_X$ to $Q = 2$ GeV for $K^0$-$\overline{K}^0$ system \cite{Ciuchini:1998ix}, to $Q = 4.6$ GeV for $B_{d,s}^0$-$\overline{B}_{d,s}^0$ system \cite{Becirevic:2001jj}, and to $Q = 2.8$ GeV for $D^0$-$\overline{D}^0$ system \cite{UTfit:2007eik}. It is seen that the running induces non-vanishing values for $C^4_M$ while $\tilde{C}^{2,3}_M$ and $C^{2,3}_M$ remain zero. Following this and using the expressions in eqs. (\ref{Xf}) and (\ref{WCs}), we numerically compute the values of all the non-vanishing Wilson coefficients at their relevant low-energy scales for each point displayed in Fig. \ref{fig:chi2}. These values are then compared with the present experimental limits obtained by the UTFit collaboration \cite{UTfit:2007eik}.

\begin{figure}[t]
\centering
\subfigure{\includegraphics[width=0.48\textwidth]{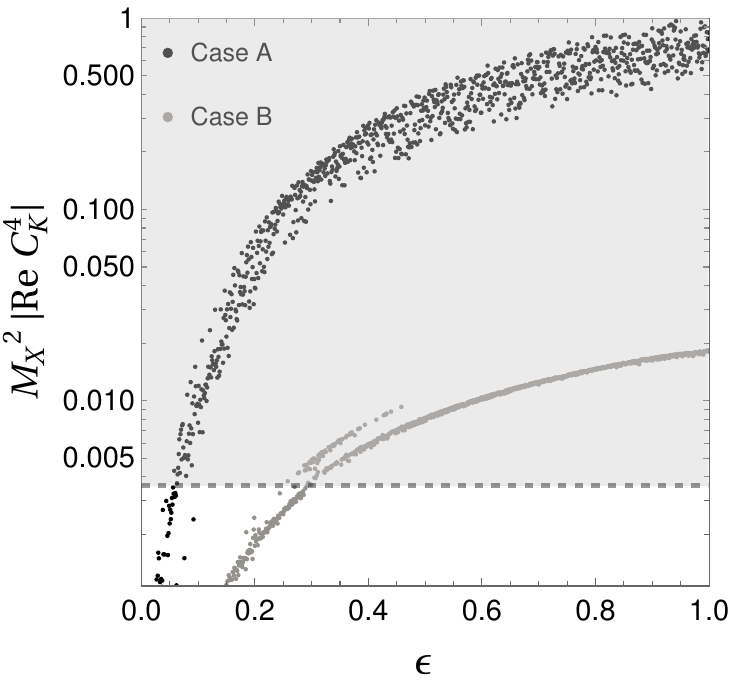}}\hspace*{0.5cm}
\subfigure{\includegraphics[width=0.48\textwidth]{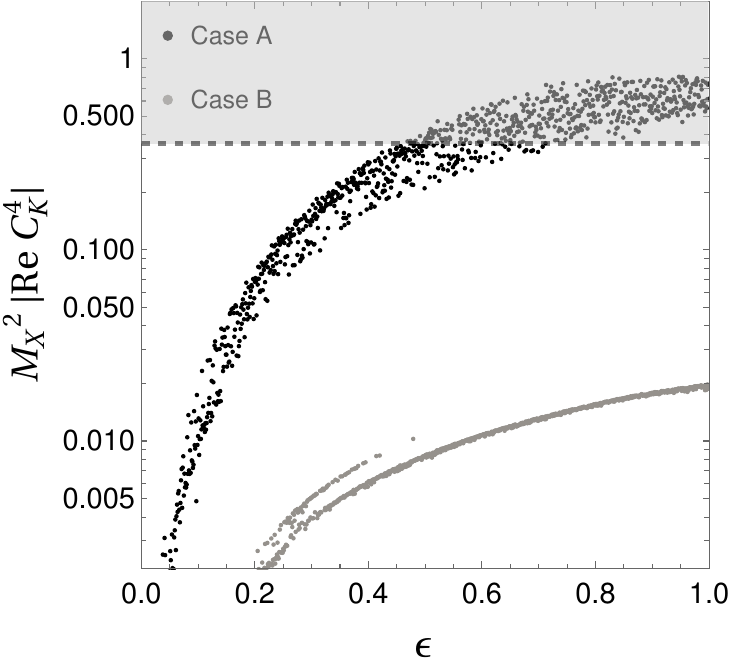}}
\caption{Magnitude of ${\rm Re}C^{4}_K$ computed for the best-fit points for different values of $\epsilon$ and for $M_X=10^6$ GeV (left panel) and $M_X=10^7$ GeV (right panel).  The shaded regions are excluded by the present limits at $95\%$ confidence level.}
\label{fig:C4}
\end{figure}

\begin{figure}[t]
\centering
\subfigure{\includegraphics[width=0.48\textwidth]{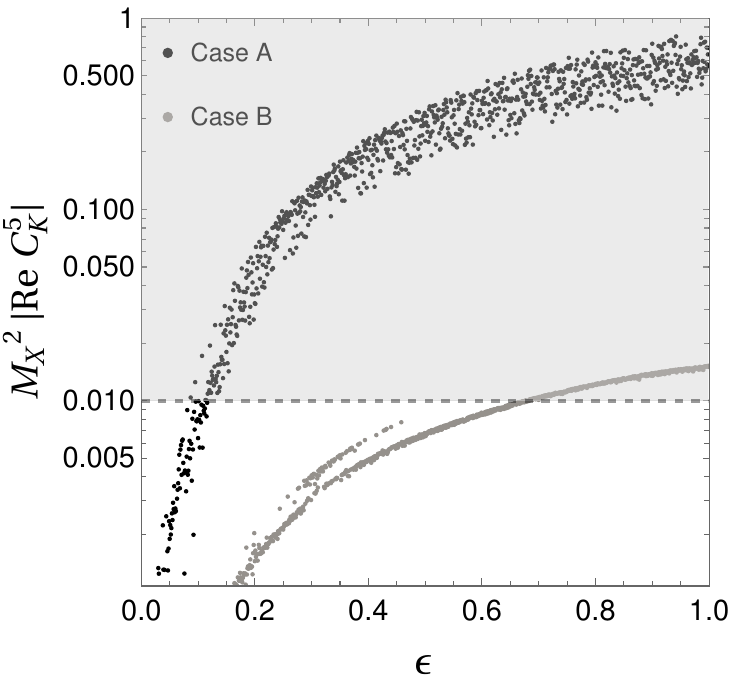}}\hspace*{0.5cm}
\subfigure{\includegraphics[width=0.48\textwidth]{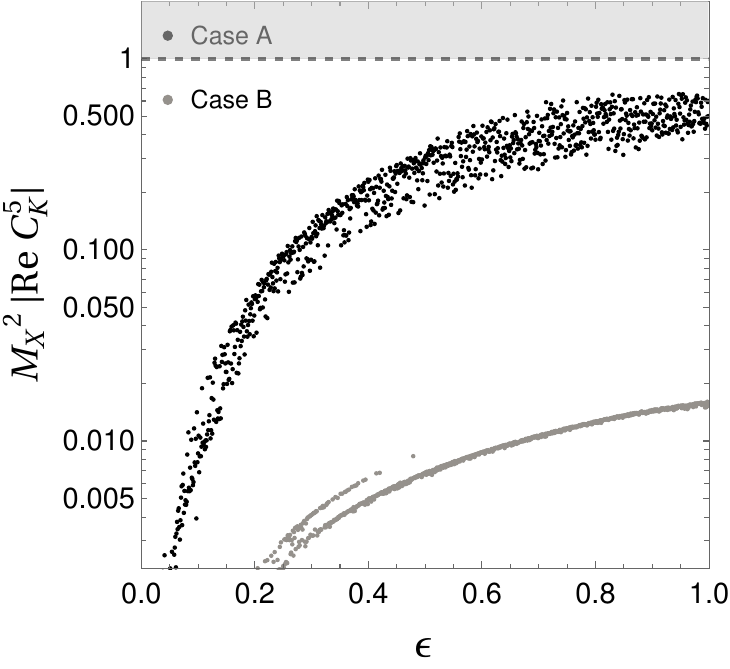}}
\caption{Magnitude of ${\rm Re}C^{5}_K$ computed for the best-fit points for different values of $\epsilon$ and for $M_X=10^6$ GeV (left panel) and $M_X=10^7$ GeV (right panel).  The shaded regions are excluded by the present limits at $95\%$ confidence level.}
\label{fig:C5}
\end{figure}

Among all the $C^i_M$ and $\tilde{C}^i_M$ computed in the present model, we find that the strongest limits on $M_X$ are almost entirely driven by  ${\rm Re} C^{4,5}_K$. To demonstrate this, we give their values as a function of $\epsilon$ for $M_X = 10^6$ and $M_X=10^7$ GeV in Figs. \ref{fig:C4} and \ref{fig:C5}, respectively. 

We also provide these estimations for cases A and B as discussed in the previous section.  As it can be verified from both the figures, the flavour constraints can be efficiently evaded for smaller values of $\epsilon$ in all the cases. Moreover, case B corresponding to strongly ordered $\mu_{di}$ and $\mu_{di}^\prime$ reduces considerably the size of $C^4_K$ and $C^5_K$. We find that the present limits on $C^4_K$ permit only $\epsilon < 0.15$ for case A and $M_X =10^6$ GeV which is already disfavoured by the large $\chi^2_{\rm min}$, see Fig. \ref{fig:chi2}. It is seen that case B for the same $M_X$ can lead to solutions with acceptable $\chi^2_{\rm min}$ and allowed ${\rm Re}C^{4,5}_K$. For $M_X=10^7$ GeV, both cases A and B can lead to viable solutions as can be read from Figs. \ref{fig:chi2}, \ref{fig:C4} and \ref{fig:C5}.

We find that the minimum mass the $X$-boson requires to evade the constraints from meson-antimeson oscillations, without relying on strongly ordered $\mu_{di}$ and $\mu_{di}^\prime$, is $M_X=3 \times 10^6$ GeV for viable spectrum of charged fermion masses and quark mixing parameters. The strong ordering can further reduce it to $M_X=10^6$ GeV. Example solutions for each are already given in Table \ref{tab:fit} and \ref{tab:sol}. For these solutions, we compute all the non-vanishing  $C^i_M$ and $\tilde{C}^i_M$ and list them in Table \ref{tab:FV}. All of them satisfy the present experimental limits. It can also be seen that the strength of Wilson coefficients corresponding to the flavour violation in the $1$-$2$ sector is typically smaller than those in the $2$-$3$ sector, as aimed by small $\epsilon$ in the present framework.
\begin{table}[t]
\begin{center}
\begin{tabular}{cccc} 
\hline
\hline
~~Flavour observables~~&~~Experimental limit~~&~~~{\bf Solution 1}~~~&~~~{\bf Solution 2}~~~\\
 \hline
Re$C_K^1$ & $[-9.6,9.6]\times 10^{-13}$     &$-5.0\times 10^{-16}$ & $-8.4\times 10^{-16}$  \\        Im$C_K^1$ & $[-9.6,9.6]\times 10^{-13}$    & $-1.7\times 10^{-30}$& $1.5\times 10^{-29}$  \\
Re$\tilde{C}_K^1$ & $[-9.6,9.6]\times 10^{-13}$ &  $-7.2\times 10^{-16}$ & ${-4.7\times 10^{-16}}$  \\
  Im$\tilde{C}_K^1$ & $[-9.6,9.6]\times 10^{-13}$  & $2.5\times 10^{-30}$ &{$3.3 \times 10^{-30}$}\\
Re$C_K^4$ & $[-3.6,3.6]\times 10^{-15}$  & $-3.2\times 10^{-15}$  & ${-3.4\times 10^{-15}}$\\ Im$C_K^4$ &  $[-1.8,0.9]\times 10^{-17}$   & $8.3\times 10^{-32}$ & {$4.1\times 10^{-29}$}\\
Re$C_K^5$ & $[-1.0,1.0]\times 10^{-14}$ &  $-2.7\times 10^{-15}$  & ${-2.8\times 10^{-15}}$\\ Im$C_K^5$ & $[-1.0,1.0]\times 10^{-14}$  & $6.8\times 10^{-32}$ &{$3.4\times 10^{-29}$}\\
\hline
$|C_{B_d}^1|$ & $<2.3\times 10^{-11}$  &  $6.6\times 10^{-18}$ & $3.3\times 10^{-18}$\\
  $ |\tilde{C}_{B_d}^1|$ & $<2.3\times 10^{-11}$   & $2.0\times 10^{-17}$ & $2.9\times 10^{-17}$\\
$|C_{B_d}^4|$ &  $<2.1\times 10^{-13}$ & $3.0\times 10^{-17}$  &  $2.5\times 10^{-17}$\\
  $|C_{B_d}^5|$ & $<6.0\times 10^{-13}$  &  $5.0\times 10^{-17}$ & $4.2 \times 10^{-17}$ \\

\hline
$|C_{B_s}^1|$ & $< 1.1 \times 10^{-9}$  &  $1.3\times 10^{-15}$ & $3.9 \times 10^{-15}$ \\
 $|\tilde{C}_{B_s}^1|$ & $< 1.1 \times 10^{-9}$  & $2.8\times 10^{-15}$ & $5.2\times 10^{-14}$\\
$|C_{B_s}^4|$ & $< 1.6 \times 10^{-11}$  &   $5.0\times 10^{-15}$ & $3.7 \times 10^{-14}$\\
 $|C_{B_s}^5|$ & $< 4.5 \times 10^{-11}$  & $8.1\times 10^{-15}$ & $6.2\times 10^{-14}$\\
\hline
$|C_D^1|$ & $<7.2 \times 10^{-13}$  &   $6.9\times 10^{-16}$  & $1.3\times 10^{-15}$ \\
 $|\tilde{C}_D^1|$ & $<7.2 \times 10^{-13}$  & $6.9\times 10^{-16}$ & $3.9 \times 10^{-14}$\\
$|C_D^4|$ & $<4.8\times 10^{-14}$   &   $2.8\times 10^{-15}$  & $2.8\times 10^{-14}$ \\
 $|C_D^5|$ & $<4.8 \times 10^{-13}$  &  $3.0\times 10^{-15}$ & $3.1\times 10^{-14}$\\
\hline
 ${\rm BR}[\mu \to e]$ & $< 7.0 \times 10^{-13}$  ~~&~$5.1 \times 10^{-17}$ & $3.2 \times 10^{-15}$\\
\hline
${\rm BR}[\mu \to 3e]$ & $< 1.0 \times 10^{-12}$   & $2.9\times 10^{-19}$ & $2.0 \times 10^{-17}$\\ 
${\rm BR}[\tau \to 3\mu]$ & $< 2.1 \times 10^{-8}$  & $5.3\times 10^{-19}$ & $2.8\times 10^{-17}$\\ 
${\rm BR}[\tau \to 3 e]$ & $< 2.7 \times 10^{-8}$  & $6.4\times 10^{-22}$ & $7.2\times 10^{-20}$ \\
\hline 
${\rm BR}[\mu \to e \gamma]$ & $< 4.2 \times 10^{-13}$  &  $6.1\times 10^{-21}$ & $2.9\times 10^{-19}$\\
${\rm BR}[\tau \to \mu \gamma]$ &  $< 4.4 \times 10^{-8}$  & $6.1\times 10^{-22}$ & $2.4\times 10^{-19}$\\
 ${\rm BR}[\tau \to e \gamma]$ &  $< 3.3 \times 10^{-8}$  & $7.3\times 10^{-24}$ & $1.9\times 10^{-22}$\\
\hline
\hline
\end{tabular}
\end{center}
\caption{The magnitudes of the various Wilson coefficients (in GeV$^{-2}$ unit) for the $\Delta F=2$ processes in the quark sector and branching ratios for the lepton flavour violating process calculated for solutions 1 and 2. The corresponding experimental limits are also listed.}
\label{tab:FV}
\end{table}

\subsection{$\mu$ to $e$ conversion}
The $X$-boson, through its flavour-conserving couplings with $u$ and $d$ quarks and flavour violating couplings with $e$ and $\mu$ leptons, can induce $\mu$ to $e$ conversion within nuclei at tree-level. The branching ratio for such a transition in the nucleus field is computed in \cite{Kitano:2002mt}. It is given by
\be \label{mu2e}
{\rm BR}[\mu \to e] = \frac{2 G_F^2}{\omega_{\rm capt}}\,(V^{(p)})^2\, \left(|g^{(p)}_{LV}|^2 + |g^{(p)}_{RV}|^2\right)\,, \ee
where the dimensionless couplings are given by
\be \label{gLV}
g^{(p)}_{LV,RV} = 2 g^{(u)}_{LV,RV} + g^{(d)}_{LV,RV}\,,\ee
with $g^{(u,d)}_{LV,RV}$ are to be identified as 
\beqa \label{gLV_Z1}
g^{(u,d)}_{LV} &\simeq& \frac{\sqrt{2}}{G_F} \frac{g_X^2}{M_X^2}\,\left(X^e_{L}\right)_{12}\, \frac{1}{2}\left[\left(X^{u,d}_L\right)_{11} + \left(X^{u,d}_R\right)_{11} \right]\,, \nonumber \\
g^{(u,d)}_{RV} &\simeq& \frac{\sqrt{2}}{G_F} \frac{g_X^2}{M_X^2}\,\left(X^e_R\right)_{12}\, \frac{1}{2}\left[\left(X^{u,d}_L\right)_{11} + \left(X^{u,d}_R\right)_{11} \right]\,,\eeqa
in the underlying model \cite{Smolkovic:2019jow}. The functions  $V^{(p)}$ and $\omega_{\rm capt}$ in eq. (\ref{mu2e}) are integral involving proton distribution in the nucleus and muon capture rate by a given nucleus, respectively.

\begin{figure}[t]
\centering
\subfigure{\includegraphics[width=0.48\textwidth]{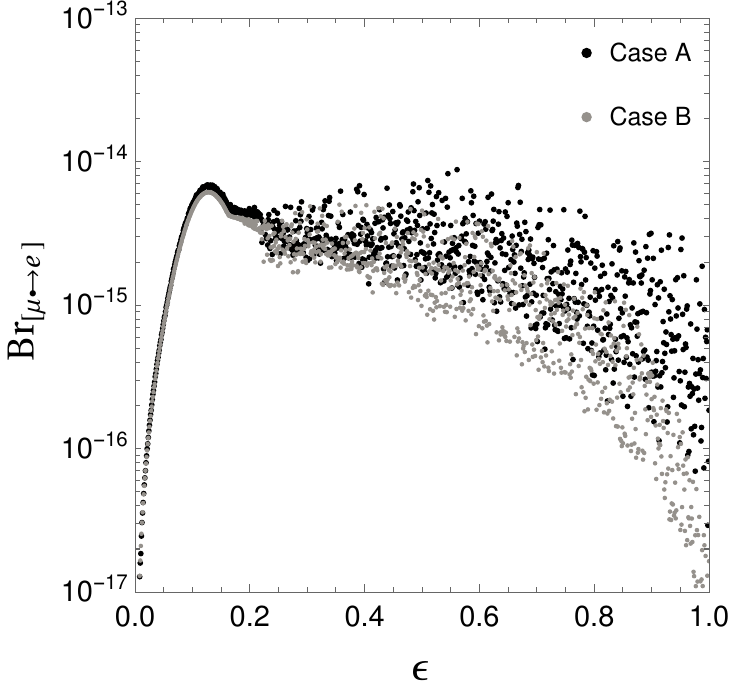}}\hspace*{0.5cm}
\subfigure{\includegraphics[width=0.48\textwidth]{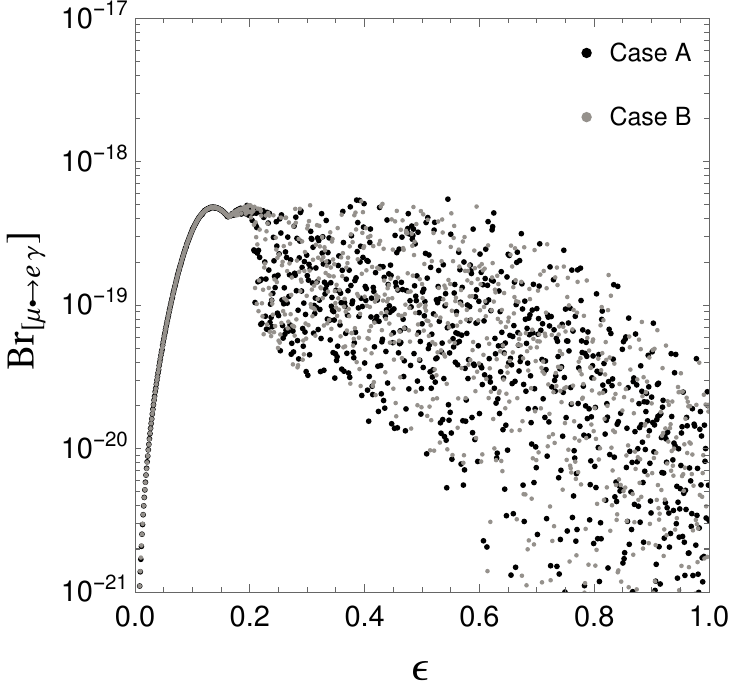}}
\caption{Left panel: ${\rm BR}[\mu \to e]$ in $^{197}$Au nucleus for various values of $\epsilon$ and for $M_X=10^6$ GeV estimated from the best-fit solutions. Right panel: The same for the lepton flavour violating observable ${\rm BR}[\mu \to e\, \gamma]$.}
\label{fig:mu2e}
\end{figure}

The SINDRUM II experiment \cite{SINDRUMII:2006dvw} has obtained the most stringent limit on $\mu$-$e$ conversion, which uses $^{197}$Au nucleus. For the latter, $V^{(p)}= 0.0974\, m_\mu^{5/2}$ and $\omega_{\rm capt} = 13.07 \times 10^{6}\,{\rm s}^{-1}$ are estimated \cite{Kitano:2002mt}. Using these values and substituting eqs. (\ref{gLV},\ref{gLV_Z1}) in eq. (\ref{mu2e}), we estimate ${\rm BR}[\mu \to e]$ for each point displayed in Fig. \ref{fig:chi2} corresponding to $M_X=10^6$ GeV. The results are displayed in the left panel of Fig. \ref{fig:mu2e}. We also estimate the same for the two benchmark solutions and list them in Table \ref{tab:FV}.

It can be seen that both cases A and B predict similar magnitude for  ${\rm BR}[\mu \to e]$. This is expected as the manipulation that differentiates between these two does not affect the flavour violating couplings in the lepton sector. $\epsilon > 0.15$ which leads to viable solutions in both cases implies ${\rm BR}[\mu \to e] \le 7 \times 10^{-15}$ which is two orders of magnitude smaller from the present limit from SINDRUM II. Therefore, the present framework does not seem to be constrained from $\mu$ to $e$ conversion.

\subsection{$l_i \to 3 l_j$ and $l_i \to l_j\, \gamma $}
The other flavour violating processes in the lepton sector include $l_i \to 3 l_j$ and $l_i \to l_j\, \gamma$. The first is mediated by the $X$ boson at tree-level in the present scenario, while the latter arises at one loop. The tri-lepton decays can be estimated as \cite{Heeck:2016xkh,Smolkovic:2019jow}
\beqa \label{lto3l}
\Gamma[l_i \to 3 l_j] &\simeq& \frac{g_X^4 m_{l_i}^5}{768 \pi^3 M_{X}^4}\,\left[ 4 {\rm Re}\left( \left(X_{eV}\right)_{ji} \left(X_{eA}\right)_{ji} \left(X_{eV}\right)^*_{jj} \left(X_{eA}\right)^*_{jj} \right) \right. \nonumber \\
&+& \left. 3 \left( \left| \left(X_{eV}\right)_{ji}\right|^2 + \left| \left(X_{eA}\right)_{ji}\right|^2 \right) \left( \left| \left(X_{eV}\right)_{jj}\right|^2 + \left| \left(X_{eA}\right)_{jj}\right|^2 \right) \right]\,,\eeqa
where, $X_{eV}$ and $X_{eA}$ are the vector and axial-vector couplings of the new gauge boson, respectively. In terms of the current in eq. (\ref{JX_model_phys}), they are defined as:
\be \label{X_VA}
X_{eV,eA} = \frac{1}{2} \left(X^e_L  \pm X^e_R\right)\,.\ee

Similarly, to compute the branching ratios for the radiative decays, $l_i \to l_j \gamma$, we use  \cite{Lavoura:2003xp}:
\be \label{muegamma}
\Gamma[l_i \to l_j \gamma] = \frac{\alpha g_X^4}{4 \pi}\,\left(1-\frac{m_{l_j}^2}{m_{l_i}^2}\right)^3\,\frac{m_{l_i}^4}{M_{X}^4}\,m_{l_i}\,\left( |c_L^\gamma|^2 + |c_R^\gamma|^2 \right)\,.\ee
Here, $\alpha$ is the fine-structure constant. The coefficients $c_L^\gamma$ and $c_R^\gamma$ are matched to the flavour violating couplings as
\beqa \label{c_gamma}
c_L^\gamma & = & \sum_{k} Q_k \left[\left(X^e_{R}\right)^*_{jk} \left(X^{e}_{R}\right)_{ik} y_{RR} +\left(X^e_{L}\right)^*_{jk} \left(X^e_{L}\right)_{ik} y_{LL} \right. \nonumber \\
&+& \left. \left(X^e_{R}\right)^*_{jk} \left(X^e_{L}\right)_{ik} y_{RL} +\left(X^e_{L}\right)^*_{jk} \left(X^e_{R}\right)_{ik} y_{LR} \right]\,.\eeqa
Similarly, $c_R^\gamma$ can be obtained by swapping $L$ and $R$ in the coupling matrices. $Q_k$ denotes the electric charge of lepton $l_k$. The loop functions $y_{LL}$, $y_{RR}$, $y_{LR}$, and $y_{RL}$ are given in \cite{Lavoura:2003xp}.

Using eqs. (\ref{lto3l}) and (\ref{muegamma}), we can estimate the branching ratios of $\mu \to 3 e$, $\tau \to 3 \mu$, $\tau \to 3 e$, $\mu \to e \gamma$, $\tau \to \mu \gamma$ and $\tau \to e \gamma$ in the present model. We find that, for $M_X \ge 10^6$ GeV, none of them is sizable enough to provide any meaningful constraint on the underlying model. Their estimated values for the benchmark solutions are provided in Table \ref{tab:FV}. It can be noticed again that due to the optimum arrangement of the flavour violating couplings, the flavour violation in the $1$-$2$ sector is typically smaller or of the same size as that in the $2$-$3$ sector. This can be considered as an improvement over the previous setups \cite{Mohanta:2022seo,Mohanta:2023soi} in which the trend is the opposite. We also give evaluated values of ${\rm BR}[\mu \to e \gamma]$ for $M_X =10^6$ GeV and several values of $\epsilon$ in Fig. \ref{fig:mu2e}.

In summary, the most stringent phenomenological constraint on the model comes from the $K^0$-$\overline{K}^0$ oscillation, which requires the new gauge boson mass at least $10^3$ TeV or higher. As discussed in the previous section, the vectorlike quarks and leptons also tend to remain closer to this scale. In this scenario, all the other phenomenological constraints such as those from the direct searches and electroweak precision tests can be trivially satisfied. The reader may refer to \cite{Mohanta:2022seo} for a detailed discussion on this.

\section{Neutrino masses}
\label{sec:neutrino}
Although the primary focus of the study has been the charged fermion mass spectrum, we elucidate how the underlying framework can be extended to accommodate the neutrino masses. The main observational features \cite{Capozzi:2017ipn,deSalas:2020pgw,Esteban:2020cvm} that make the latter different are: (a) the overall mass scale of neutrinos is several orders of magnitude smaller than that of the charged fermions, and (b) neutrinos are relatively less hierarchical since $m_{\nu 2}/m_{\nu 3} \simeq 0.2$ for $m_{\nu 1} = 0$ and $m_{\nu 2}/m_{\nu 3} \simeq 1$ for the quasi-degenerate neutrinos. As we outline below, the present framework can be extended in two qualitatively different ways to account for the neutrino masses with the aforementioned features. They are along the same lines of the usual SM extensions for the neutrino masses.

\subsection{Majorana option}
Neutrinos can be of Majorona nature, leading to a presence of the lepton number violation in the framework. Assuming that the scale characterizing this, namely $\Lambda_{\rm LN}$, is greater than the electroweak as well as $U(1)_F$ breaking scale, the neutrino masses can be best described by the usual dimension-5 operator \cite{Weinberg:1979sa}. In the present framework, such operators are obtained as
\be \label{dim5}
{\cal L}_{\rm dim-5} \supset \frac{c^{(1)}_{ij}}{2 \Lambda_{\rm LN}}\, \left(\overline{L}_{Li} H_{u i}\right)\left(H^T_{u j} L^c_{Lj}\right)\,+\,\frac{c^{(2)}_{ij}}{2 \Lambda_{\rm LN}}\, \left(\overline{L}_{Li} H_{u j}\right)\left(H^T_{u i} L^c_{Lj}\right)\,+\,{\rm h.c.}\,.\ee
The first operator can be obtained from a full theory by integrating out fermions neutral under the SM and $U(1)_F$ gauge symmetries. This is identical to the type I seesaw mechanism \cite{Minkowski:1977sc,Yanagida:1979as,Mohapatra:1979ia,Schechter:1980gr}. The ultraviolet completion for the second operator requires either $U(1)_F$ neutral fermions charged under the electroweak symmetry (type III seesaw \cite{Foot:1988aq}) or complex scalars charged under both the electroweak and new gauge symmetry (type II seesaw \cite{Lazarides:1980nt,Magg:1980ut,Mohapatra:1980yp}). It is straightforward to verify that none of these extensions leads to additional contributions to the anomalies.

Consequent upon the electroweak symmetry breaking, the neutrino masses are given by
\be \label{mnu_maj}
\left(m_\nu\right)_{ij} = \frac{c_{ij}}{\Lambda_{\rm LN}}\, \mu_{ui}\,\mu_{uj}\,,\ee
where $c_{ij}$ represent coefficients appearing in eq. (\ref{dim5}) scaled appropriately by the Yukawa couplings $y_{ui}$. With the most general $c_{ij}$, all the three neutrinos obtain tree-level masses, suppressed by $\Lambda_{\rm LN}$. This is following both the features (a) and (b) mentioned at the beginning of this section. Moreover, the coefficients also provide enough freedom to reproduce the viable leptonic mixing parameters. A more predictive and insightful approach to the latter would require additional symmetries and/or structure for the specific UV completion, and it is a subject of elaborate model building.

\subsection{Dirac option}
Alternatively, the neutrino masses can also be introduced in an identical way it is done for the charged fermions in the framework. This minimally requires three Weyl fermions, $\nu_{Ri}$, with $U(1)_F$ charges $(1-\epsilon, 1+\epsilon, -2)$ and a neutral vectorlike pair $N_{L,R}$. The former is already present in the original model from an anomaly cancellation requirement, see Table \ref{tab:fields}. All of these are singlet under the SM gauge symmetries. At the leading order, the Dirac neutrino mass matrix is obtained as 
\be \label{M0_eff_f_neut}
\left(M^{(0)}_\nu\right)_{ij} = -\frac{1}{m_N}\, \mu_{\nu i}\, \mu_{\nu j}^\prime\,,\ee
where $\mu_{\nu i} = y_{\nu i} \langle H_{u i} \rangle$, $\mu^\prime_{\nu i} = y^\prime_{\nu i} \langle \eta^*_i \rangle$ and $m_N$ is the Dirac mass for the pair $N_{L,R}$. The above leads to one massive light neutrino state.

Because of the universal seesaw-like structure present in this framework, the smallness of the neutrino mass can be attributed to large $m_N$. Identifying the massive neutrino state with the atmospheric neutrino oscillation scale \cite{Esteban:2020cvm}, one finds
\be \label{mN_est}
m_N \approx 2\times 10^{17}\, {\rm GeV}\,\left(\frac{0.05\,{\rm eV}}{m_{\nu 3}}\right)  \left(\frac{\langle \eta \rangle}{100\,{\rm TeV}}\right)  \left(\frac{\langle H_u \rangle}{100\,{\rm GeV}}\right)\,,\ee
where we have considered all the dimensionless parameters of order one. Apparently, the issue of unnaturally small neutrino Yukawa couplings in the ordinary extension of the SM with Dirac neutrinos transforms into the issue of large hierarchy, $m_N \gg m_{U,D,E}$, in the present framework.

The solar neutrino mass scale can arise in the setup when the higher order corrections are introduced to the $M^{(0)}_\nu$. The same expression, eq. (\ref{M2_eff_fnl_f}), would apply in this case with appropriate changes. However, some tuning may be required as the desired magnitude of $m_{\nu 2}/m_{\nu 3}$ is larger than the typical loop suppression factor. It is also possible to avoid this by introducing two or more copies of $N_{L,R}$ and multiplets of $\nu_{Ri}$ such that the tree-level neutrino mass matrix has rank more than one. In this case, both the solar and atmospheric scales are generated at the tree level and are expected to be not-so-hierarchical.

\section{Summary and further issues}
\label{sec:summary}
What distinguishes the framework of radiative fermion masses from other models of flavour hierarchies is its ability to provide at least some of the fermion masses as calculable quantities. Here, calculability means that these masses can be fully or partially expressed in terms of quantities that can be measured independently in experiments, at least in principle. Such quantities must arise from new physics interactions, as the standard model alone cannot accommodate the radiative mass mechanism. This opens up an avenue for model building aimed at constructing viable and predictive models for radiative mass generation. In the present work, we systematically explore the new gauge sector that can enable this.

It is shown that the SM extended with just a single abelian gauge symmetry is sufficient to provide a suitable framework for the radiative induction of the masses of the lighter generations. The Yukawa sector at the tree level can be engineered to possess a global $U(2)^5$ symmetry corresponding to the massless first and second generations. This symmetry does not commute with the flavour-dependent gauge symmetry, and it gets broken completely if all three generations of the SM fermions are charged differently under the latter. This then generates the masses of the second generation at one loop and of the first generation at two loops. The 2-loop corrected mass matrix can be completely expressed in terms of tree-level mass matrix and the gauge charges and gauge boson mass, as shown in eq. (\ref{M2_eff_fnl}). This offers significant advancements over our previous attempts in this direction, in which the masses of both the lighter generations were induced at one-loop and inter-generational hierarchy was attributed to specific ordering in the gauge boson masses. These models also required enlarged gauge sector in comparison to the one presented here.

Large flavour changing neutral currents mediated by the gauge boson of new gauge symmetry are inherently present in this kind of framework. They put the strongest constraint on the scale of new physics. As we have shown in this study, the flavour changing couplings can be ordered in such a way that their strengths in the $1$-$2$ sector are the smallest. This is phenomenologically favourable since the most stringent constraint on flavour violations comes from the $1$-$2$ sector, in particular, from $K^0$-$\overline{K}^0$ mixing and $\mu$-$e$ conversion in nuclei. The smallness of flavour changing couplings in $1$-$2$ is proportional to the difference between the charges of the first and second-generation fermions under the new gauge symmetry. The latter cannot be made arbitrarily small, as complete degeneracy in these charges leads to the strictly massless first generation. This correlation is then explored to find the optimum strength of flavour violation through a comprehensive numerical analysis, and we find that the down-type quark sector predominantly governs this optimization. If the down quark mass is to be reproduced within the $3 \sigma$ range of its present value extracted from the lattice computation, then the mass of the $U(1)_F$ gauge boson can not be lower than $10^3$ TeV provided that all the flavour constraints are satisfied. This lower limit is smaller by nearly two orders of magnitude in comparison to our previous frameworks proposed along the same lines \cite{Mohanta:2022seo,Mohanta:2023soi}.

It is remarkable that extending the SM with a single abelian gauge symmetry can enable its gauge sector to support radiative mass generation for lighter fermions. However, there are areas where this framework requires further improvement. For example,
\begin{itemize}
\item Calculability is not fully achieved because the Yukawa sector of the theory still contains numerous couplings. These couplings may be reduced if $U(1)_F$ is replaced with a gauge symmetry under which the three generations transform as an irreducible representation (see, for example, \cite{Mohanta:2023soi}). However, this would enrich the gauge sector, and it remains to be seen whether the salient features of the minimal gauge sector presented here can be retained. Another option to reduce Yukawa couplings to some extent is to implement this scheme in models that provide quark-lepton unification \cite{Georgi:1974sy,Fritzsch:1974nn,Pati:1974yy} at high energies.
\item Although the framework accounts for the hierarchical nature of charged fermion masses, it does not explain the smallness of mixing in the quark sector. This is because the underlying symmetries and the field content of the model allow different and arbitrary tree-level Yukawa couplings for the up-type and down-type quarks. Frameworks with an improved degree of calculability can address this issue.
\item The lowest scale of new physics remains at $10^3$ TeV. While this scale, or even a larger one, does not hinder the core mechanism discussed here, the significant separation between it and the electroweak scale poses a challenge. This separation can be managed at the expense of fine-tuning in the generic case. However, in more ambitious theories where the parameters of the scalar potential are also calculable, this becomes a concrete problem.
\end{itemize}
We believe that the above issues require a set of systematic investigations to establish further the validity and robustness of the radiative mass generation scheme.

\section*{Acknowledgements}
This work is supported by the Department of Space (DOS), Government of India. KMP acknowledges partial support under the MATRICS project (MTR/2021/000049) from the Science \& Engineering Research Board (SERB), Department of Science and Technology (DST), Government of India.

\appendix
\section{On the origin of the gauge charges}
\label{app:KM}
A choice made in eq. (\ref{gauge_charges}) for the charges of underlying abelian flavour symmetry can simply be obtained from kinetic mixing. Consider two $U(1)$ symmetries with gauge bosons $X_{1,2}^\mu$ and the gauge interactions
\be \label{LG_app}
- {\cal L}_G = g^{(1)}\, q^{(1)}_{ii}\, \overline{f}^\prime_{i}\,\gamma^\mu\,f^\prime_{i}\, X^{(1)}_\mu + g^{(2)}\, q^{(2)}_{ii}\, \overline{f}^\prime_{i}\,\gamma^\mu\,f^\prime_{i}\, X^{(2)}_{\mu}\,,  \ee
where $f^\prime$ stands for both $f^\prime_L$ and $f^\prime_R$. An appropriate choice for the charge matrices is
\be \label{q1-q2}
q^{(1)} = {\rm Diag.}(1,1,-2)\,,~~q^{(2)} = {\rm Diag.}(1,-1,0)\,.\ee

Next, consider the kinetic terms of these gauge bosons with non-vanishing kinetic mixing \cite{Holdom:1985ag}
\be \label{kin}
-{\cal L}_{\rm kin} = \frac{1}{4} F^{(1)}_{\mu \nu} F^{(1) \mu \nu} + \frac{1}{4} F^{(2)}_{\mu \nu} F^{(2) \mu \nu} + \frac{\chi}{2} F^{(1)}_{\mu \nu} F^{(2) \mu \nu}\,,\ee
with $F^{(\alpha)}_{\mu \nu}$ denote the field strength of the gauge boson $X^{(\alpha)}_\mu$. The kinetic terms can be diagonalized by a linear transformation
\be \label{LT_kin}
\left(\ba{c} X^{(1)}_\mu \\ X^{(2)}_\mu \ea \right) \to \left(\ba{cc} 1 & 0  \\ -\chi & 1 \ea \right)  \left(\ba{c} X^{(1)}_\mu \\ X^{(2)}_\mu \ea \right)\,.  \ee
This removes the kinetic mixing from the kinetic terms, which then reappear in the gauge interactions. Substituting (\ref{LT_kin}) in eq. (\ref{LG_app}) leads to
\be \label{LG_app_1}
- {\cal L}_G = g^{(1)} \left(q^{(1)}_{i i} - \chi \frac{g^{(2)}}{g^{(1)}}\, q^{(2)}_{i i} \right)\, \overline{f}^\prime_{i}\,\gamma^\mu\,f^\prime_{i}\, X^{(1)}_{\mu} + g^{(2)} q^{(2)}_{i i}\, \overline{f}^\prime_{i}\,\gamma^\mu\,f^\prime_{i}\, X^{(2)}_{\mu}\,.  \ee
Setting $g^{(1)}=g_X$, $\chi g^{(2)}/g^{(1)} \equiv \epsilon$ and $X^{(1)}_\mu = X_\mu$, one obtains the desired charges given in eq. (\ref{gauge_charges}).  By setting $M_{X_2} \gg M_{X}$, the effects of the gauge interactions associated with $X^{(2)}_\mu$ gauge boson can be decoupled from the effective theory.

The two $U(1)$ symmetries with flavour non-universal charges in eq. (\ref{q1-q2}) along with any flavour universal $U(1)$ (for example hypercharge in the present framework) can be recast into $U(1)_1 \times U(1)_2 \times U(1)_3$ such that only $i^{\rm th}$ generation is charged under $U(1)_i$. To see this, consider gauge interactions:
\be \label{three_u1}
\sum_{\alpha = 1}^3\,g^{(\alpha)}\, q^{(\alpha)}_{ii}\, \overline{f}^\prime_{i}\,\gamma^\mu\,f^\prime_{i}\, X^{(\alpha)}_{\mu}\,,  \ee
with $q^{(1),(2)}$ as already given in eq. (\ref{q1-q2}) and $q^{(3)} = {\rm Diag.}(1,1,1)$. A rotation 
\be \label{rot_gb}
X_\mu^{(\alpha)} \to \tilde{X}_\mu^{(\alpha)} = {\cal R}_{\alpha \beta}\,X_\mu^{(\beta)}\,,\ee
can be performed to redefine the gauge bosons such that in the new basis, the gauge interactions become 
\be \label{three_u1_2}
\sum_{\alpha = 1}^3\,\tilde{g}^{(\alpha)}\, \tilde{q}^{(\alpha)}_{ii}\, \overline{f}^\prime_{i}\,\gamma^\mu\,f^\prime_{i}\, \tilde{X}^{(\alpha)}_{\mu}\,.  \ee
The new couplings are then given by
\be \label{rot_gc}
\tilde{g}^{(\alpha)} \tilde{q}_{ii}^{(\alpha)} = {\cal R}_{\alpha \beta}\,g^{(\beta)} q_{ii}^{(\beta)} \,,\ee
where ${\cal R}$ is an orthogonal matrix.

For a specific choice, $\sqrt{6} g^{(1)}=\sqrt{2} g^{(2)}=\sqrt{3} g^{(3)} \equiv \tilde{g}$, and 
\be \label{R_gb}
{\cal R} = \left(
\begin{array}{ccc}
 \frac{1}{\sqrt{6}} & \frac{1}{\sqrt{2}} & \frac{1}{\sqrt{3}} \\
 \frac{1}{\sqrt{6}} & -\frac{1}{\sqrt{2}} & \frac{1}{\sqrt{3}} \\
 -\sqrt{\frac{2}{3}} & 0 & \frac{1}{\sqrt{3}} \\
\end{array}
\right)\,,\ee
one finds,
\be \label{q_tilde}
 \tilde{g}^{(1)} \tilde{q}^{(1)} = \tilde{g}\,{\rm Diag.}(1,0,0)\,,~~\tilde{g}^{(2)} \tilde{q}^{(2)} = \tilde{g}\,{\rm Diag.}(0,1,0)\,,~~\tilde{g}^{(3)}\tilde{q}^{(3)} = \tilde{g}\,{\rm Diag.}(0,0,1)\,.\ee
Therefore, the three abelian symmetries of the underlying theory, i.e. two flavour non-universal new abelian symmetries and a hypercharge, can be put in a form such that each generation is exclusively charged under only one $U(1)$ with equal strength of interactions. Recently, this kind of framework has been put forward under the name of ``tri-hypercharge'' \cite{FernandezNavarro:2023rhv} or ``deconstructed hypercharge'' \cite{Davighi:2023evx}.

\section{Scalar sector}
\label{app:scalar}
In this Appendix, we discuss the scalar sector of the framework and outline its phenomenological implications. The model contains total nine scalar multiplets, $H_{u i}$, $H_{d_i}$ and $\eta_i$ with $i=1,2,3$.  The SM and $U(1)_F$ invariant renormalisable scalar potential can be written as:
\beqa \label{potential}
V &=& m_{u i}^2\, H_{u i}^\dagger H_{u i}\, +\, m_{d i}^2\, H_{d i}^\dagger H_{d i}\, +\, m_{\eta i}^2\, \eta_{i}^\dagger \eta_{i}\, \nonumber \\
& + & \left\{(m_{ud\eta})_{ijk}\, \epsilon_{ijk}\, \eta_i H_{u j} H_{d k}\,+\, (m_{\eta})_{ijk}\, \epsilon_{ijk}\, \eta_i \eta_j \eta_k + {\rm h.c.}\right\}\, \nonumber \\
& + &  (\lambda_u)_{ij}\, H_{u i}^\dagger H_{u i} H_{u j}^\dagger H_{u j}\,+\, (\lambda_d)_{ij}\, H_{d i}^\dagger H_{d i} H_{d j}^\dagger H_{d j}\,+\, (\lambda_\eta)_{ij}\, \eta_{i}^\dagger \eta_{i} \eta_{j}^\dagger \eta_{j}\, \nonumber \\
& + &  (\lambda_{ud})_{ij}\, H_{u i}^\dagger H_{u i} H_{d j}^\dagger H_{d j}\,+\, (\lambda_{u\eta})_{ij}\, H_{u i}^\dagger H_{u i} \eta_j^\dagger \eta_j\,+\, (\lambda_{d \eta})_{ij}\, H_{d i}^\dagger H_{d i} \eta_{j}^\dagger \eta_{j}\, \nonumber \\
& + &  (\tilde{\lambda}_{ud})_{ij}\, H_{u i}^\dagger H_{u j} H_{d j}^\dagger H_{d i}\,+\, (\tilde{\lambda}_{u\eta})_{ij}\, H_{u i}^\dagger H_{u j} \eta_j^\dagger \eta_i\,+\, (\tilde{\lambda}_{d \eta})_{ij}\, H_{d i}^\dagger H_{d j} \eta_{j}^\dagger \eta_{i}\, \nonumber \\
& + & \left\{(\lambda_{ud\eta})_{ij}\, \eta_i^\dagger H_{ui} \eta_j^\dagger H_{dj} + (\tilde{\lambda}_{ud\eta})_{ij}\, \eta_i^\dagger H_{uj} \eta_j^\dagger H_{di} + {\rm h.c.}\right\}\,.\eeqa 
Note that the above potential is identical to the one obtained in our previous framework based on two flavour non-universal abelian symmetries \cite{Mohanta:2022seo}. Consequent upon the gauge symmetry breaking this leads to several electrically charged and neutral physical scalars. In principle, these states can also contribute to the charged fermion masses through loops. However, such contributions can remain relatively suppressed under reasonable conditions as we discuss below.

Identifying the electromagnetically neutral scalars residing in $H_{ui}$, $H_{di}$ and $\eta_i$ as $h_{ui}$, $h_{d i}$ and $\eta_i$, respectively, their mass term can be parametrized as 
\be \label{hmass}
\frac{1}{2} \left(M^2_h \right)_{ab}\,\tilde{h}_a\,\tilde{h}_b\,, \ee
where $a=1,...,9$ and $\tilde{h} = (h_{ui},h_{dj},\eta_k)^T$. The $9 \times 9$ symmetric neutral scalar mass matrix $M^2_h$ can be written in terms of $3 \times 3$ blocks as
\be \label{M9}
M^2_h = \left(\ba{ccc} m_{u u}^2 & m^2_{ud} &  m^2_{u\eta} \\ (m^2_{ud})^T & m_{d d}^2 & m^2_{d\eta}\\ (m^2_{u\eta})^T & (m^2_{d\eta})^T & m^2_{\eta \eta} \ea \right)\,.\ee
The above parameters can be computed from the scalar potential and they are functions of various parameters appearing in eq. (\ref{potential}) and VEVs of $H_{ui}$, $H_{di}$ and $\eta_i$. As usual, the physical neutral scalar states $h_a$ can be obtained by diagonalizing $M^2_h$. Explicitly, we define
\be \label{R}
\tilde{h}_a = (R_h)_{ab}\, h_b\,,\ee
where $R_h$ is orthogonal matrix such that $R_h^T M^2_h R_h = {\rm Diag.}(m_{h_1}^2,...,m^2_{h_9})$. 
 Although the typical mass scale of the scalars is ${\cal O}(M_X)$, one of them must be identified with the observed SM-like Higgs boson with a mass of 125 GeV \cite{ATLAS:2015yey}. Minimally, the lightest of the neutral scalars, $h_1$                                                                                                                                                                                                                                                                                                                                                             , can play this role. This requires ${\rm Det.} M_h^2 \ll M_X^{18}$, necessitating a tuning of the parameters in the scalar potential. Given the large number of incalculable parameters in the scalar potential, we assume that such an arrangement is possible. As noted in section \ref{sec:summary}, this becomes a concrete problem in more ambitious frameworks where the scalar potential is better constrained.

Let us first comment on the loop-induced contributions of these scalars to the fermion masses. Since the left-chiral fermions of the SM couple only to $H_{ui}$ or $H_{di}$, and the right-chiral fermions couple solely to $\eta_i$ at tree-level (see eq. (\ref{LY_SM})), the one-loop diagrams mediated by the neutral scalars depend on the mixing between $H_{ui,di}$ and $\eta_i$. These  contributions can be small if $m^2_{\eta \eta} \gg m^2_{u\eta}, m^2_{d\eta}$ in eq. (\ref{hmass}). In this limit, the physical states $h_{1,...,6}$ ($h_{7,...,9}$) consist of dominantly $H_{ui,di}$  ($\eta_i$) with components of $\eta_i$ ($H_{ui,di}$) suppressed by a factor of ${\cal O}(m^2_{d\eta} (m^2_{\eta \eta})^{-1})$ or ${\cal O}(m^2_{u\eta} (m^2_{\eta \eta})^{-1})$. Hence, $h_a$ mediated diagrams can provide only sub-leading contributions to $\delta M^{(1,2)}$.  The same argument also applies to the charged scalars. As they arise entirely from $H_{di}$ and $H_{ui}$ with no counterparts in $\eta_i$ to mix with, they do not contribute to the fermion masses at 1-loop. Thus, the scalar-induced contributions in this setup can be made suppressed if small enough mixing between $H_{ui,di}$ and $\eta_i$ is considered.

The multiplicity of scalars and the feature that Yukawa couplings are not proportinal to the masses for the SM fermions in the underlying framework leads to flavour-changing transitions mediated by the neutral scalars. For example, consider the interactions involving the up-type quarks in eq. (\ref{LY_SM}). In the physical basis, they lead to the following flavour changing interactions between the SM up-type quarks and neutral scalars:
\beqa \label{hup}
\left(\left(U^{u \dagger}_{L } \right)_{mi} {y_u}_i (R_u)_{ia} \left(\rho_R^\dagger U^u_R \right)_n+  \left(U^{u\dagger}_{L} \rho_L \right)_m {y_u^{\prime}}_k (R_\eta)_{k a} \left(U^u_{R} \right)_{kn} \right) \overline{u}_{Lm} {h}_a u_{Rn} + {\rm h.c.}\,,   \eeqa
at the leading order. Since the masses of the heavy neutral scalars are $\gtrsim {\cal O}(10^6)$ GeV, the FCNCs mediated by them are sufficiently small. The existing experimental constraints therefore only require that the flavor-changing couplings of the light Higgs be suppressed. Given the large number of parameters in the scalar potential, we assume that such an arrangement is possible, albeit with some fine-tuning.

\section{Details of numerical analysis}
\label{app:numerical}
It can be seen from the Lagrangian in eq. (\ref{LY_SM}) that the parameters $m_U$, $m_D$ and $m_E$ can be made real by absorbing their phases into the fields $U_L$, $D_L$ and $E_L$, respectively. Similarly, rephasing of $Q_{Li}$, $L_{Li}$, $u_{Ri}$, $d_{Ri}$ and $e_{R i}$ can get rid of unphysical phases of $y_{ui}$, $y_{ei}$, $y_{ui}^\prime$, $y_{di}^\prime$ and $y_{ei}^\prime$, respectively. The parameter $y_{d3}$ can also be made real by rephasing $D_R$. Furthermore, we assume that all the VEVs are real. The gauge sector has three real parameters, $g_X$, $\epsilon$ and $M_X$, out of which $g_X$ can be fixed to some particular value without loss of generality and the remaining two parameters then control the strength of loop corrections. Hence, the model has 25 real parameters, i.e. real $\epsilon$, $M_X$, $m_U$, $m_D$, $m_E$, $\mu_{ui}$, $\mu_{ei}$, $\mu_{ui}^\prime$, $\mu_{di}^\prime$, $\mu_{ei}^\prime$, $\mu_{d3}$ and complex $\mu_{d1}$, $\mu_{d2}$. These can be used to reproduce $13$ observed quantities which include $9$ charged fermion masses, $3$ quark mixing angles and a Dirac CP phase in the quark sector.

The fitting of observables is carried out using the $\chi^2$ optimization procedure. The $\chi^2$ function is defined as 
\be \label{chi2}
 \chi^2 = \sum_{i} \left( \frac{O^i_{\rm th}-O^i_{\rm exp}}{\sigma_i}\right)^2\,,\ee
where $i=1,..,13$ represents the thirteen observables. $O^i_{\rm th}$ is the theoretically computed value of the $i^{\rm th}$ observable in terms of the input parameters listed above. We carry out this computation at the renormalization scale $Q=M_Z$ as discussed in section \ref{sec:SM}. $O^i_{\rm exp}$ ($\sigma^i$) are the corresponding mean (standard deviation) value extracted from the experiments and evolved at $Q=M_Z$. The values of charged fermion masses are taken from \cite{Xing:2007fb} and quark mixing parameters from \cite{ParticleDataGroup:2022pth}. Note that for the charged lepton masses, we use conservative $10\%$ uncertainty, as done before in \cite{Mummidi:2021anm,Mohanta:2022seo,Mohanta:2023soi}. All these values are listed in the second column in Table \ref{tab:fit} for convenience and comparison.

\begin{table}[t]
\begin{center}
\begin{tabular}{ccc} 
\hline
\hline
~~Parameters~~&~~\textbf{Solution 1} ~~&~~\textbf{Solution 2}~~\\
\hline
$M_X$           & $3\times 10^6$   & $10^6$             \\
$\epsilon$      & $ 0.178$         & $ 0.285 $        \\
\hline
$m_U$   & $5.6653\times 10^{6 }$  & $ 6.2838\times 10^{6 } $   \\
$m_D$   & $4.9336 \times 10^{8}$  & $ 1.5524\times 10^{ 8} $   \\
$m_E$   & $ 8.2489\times 10^{6}$  & $ 3.9884\times 10^{ 6} $   \\
\hline
$\mu_{u1}$ & $-9.6518$            & $ 1.2821 $          \\
$\mu_{u2}$ & $ 9.9336 $           & $ -7.2968$          \\
$\mu_{u3}$ & $ -3.7757\times10^2$ & $ 4.6534\times10^2$ \\
\hline
$\mu^\prime_{u1}$ & $6.9209\times 10^5 $    & $1.1248\times 10^{6}$   \\
$\mu^\prime_{u2}$ & $-1.6434 \times 10^6 $  & $-1.4161 \times 10^{6}$ \\
$\mu^\prime_{u3}$ & $-1.8288\times 10^6 $   & $1.4585\times 10^{6}$   \\
\hline
$\mu_{d1}$ & $-1.8003 \times 10^{1} + i\ 2.4815  $  & $1.9233 + i\ 1.9146 $           \\
$\mu_{d2}$ & $2.9553  \times 10^{1} + i\ 1.6954$    & $-3.2718+ i\ 1.9150\times 10^1$ \\
$\mu_{d3}$ & $-2.2993 \times 10^2$                  & $2.2572\times 10^2$             \\
\hline
$\mu^\prime_{d1}$ & $-7.1617\times 10^5 $ & $-4.8001\times 10^4 $ \\
$\mu^\prime_{d2}$ & $7.2368\times 10^5$   & $ 4.8498\times 10^5 $ \\
$\mu^\prime_{d3}$ & $4.4255 \times 10^6$  & $1.4044 \times 10^6 $ \\
\hline
$\mu_{e1}$     & $2.5748\times 10^1 $     & $ -1.3349 \times 10^1$   \\
$\mu_{e2}$     & $-2.6008\times 10^{1} $  & $-1.5363\times 10^1 $    \\
$\mu_{e3}$     & $-4.7027 \times10^1$     & $2.1028\times 10^1 $     \\
\hline
$\mu^\prime_{e1}$ & $-1.3925\times 10^5$    & $3.6639 \times 10^4 $   \\
$\mu^\prime_{e2}$ & $-1.4093 \times 10^5 $  & $-7.9852 \times 10^4 $  \\
$\mu^\prime_{e3}$ & $-1.4252 \times 10^5 $  & $-2.1704 \times 10^5 $  \\
\hline
\hline
\end{tabular}
\end{center}
\caption{Optimized values of various input parameters obtained for two example solutions displayed in Table \ref{tab:fit}. All the values of dimension-full parameters are in GeV.}
\label{tab:sol}
\end{table}
While performing the optimization, we also ensure that various parameters do not take unrealistic values. For example, the perturbativity limits would imply $y_{fi} \le \sqrt{4 \pi}$. Since $\langle H_{fi} \rangle$ contributes to the electroweak symmetry breaking, one expects $\langle H_{fi} \rangle < 246$ GeV. Therefore, we impose that $\mu_{f i} < \sqrt{4 \pi} \times 246$ GeV. Similarly,   $\mu^\prime_{f i} < \sqrt{4 \pi}\, M_X$ is also imposed, since these parameters break the $U(1)_F$ symmetry and hence their values cannot be significantly higher than the underlying symmetry breaking scale. The masses of vectorlike fermions do not break any symmetry, and hence they are kept unrestricted. 

The fitted values of various input parameters for two example solutions as discussed in section \ref{subsec:numerical} are displayed here in Table \ref{tab:sol}.  Both the solutions correspond to the lowest possible allowed values of $\epsilon$ for which $\chi^2_{\rm min} \lesssim 5$. 

\bibliography{biblio}

\providecommand{\href}[2]{#2}\begingroup\raggedright\begin{thebibliography}{10}

\bibitem{Weinberg:1972ws}
S.~Weinberg, \emph{{Electromagnetic and weak masses}},
  \href{https://doi.org/10.1103/PhysRevLett.29.388}{\emph{Phys. Rev. Lett.}
  {\bfseries 29} (1972) 388}.

\bibitem{Georgi:1972hy}
H.~Georgi and S.L.~Glashow, \emph{{Attempts to calculate the electron mass}},
  \href{https://doi.org/10.1103/PhysRevD.7.2457}{\emph{Phys. Rev. D} {\bfseries
  7} (1973) 2457}.

\bibitem{Mohapatra:1974wk}
R.N.~Mohapatra, \emph{{Gauge Model for Chiral Symmetry Breaking and Muon
  electron Mass Ratio}},
  \href{https://doi.org/10.1103/PhysRevD.9.3461}{\emph{Phys. Rev. D} {\bfseries
  9} (1974) 3461}.

\bibitem{Barr:1978rv}
S.M.~Barr and A.~Zee, \emph{{Calculating the Electron Mass in Terms of Measured
  Quantities}}, \href{https://doi.org/10.1103/PhysRevD.17.1854}{\emph{Phys.
  Rev. D} {\bfseries 17} (1978) 1854}.

\bibitem{Wilczek:1978xi}
F.~Wilczek and A.~Zee, \emph{{Horizontal Interaction and Weak Mixing Angles}},
  \href{https://doi.org/10.1103/PhysRevLett.42.421}{\emph{Phys. Rev. Lett.}
  {\bfseries 42} (1979) 421}.

\bibitem{Yanagida:1979gs}
T.~Yanagida, \emph{{Horizontal Symmetry and Mass of the Top Quark}},
  \href{https://doi.org/10.1103/PhysRevD.20.2986}{\emph{Phys. Rev. D}
  {\bfseries 20} (1979) 2986}.

\bibitem{Barbieri:1980tz}
R.~Barbieri and D.V.~Nanopoulos, \emph{{Hierarchical Fermion Masses From Grand
  Unification}},
  \href{https://doi.org/10.1016/0370-2693(80)90395-0}{\emph{Phys. Lett. B}
  {\bfseries 95} (1980) 43}.

\bibitem{Balakrishna:1987qd}
B.S.~Balakrishna, \emph{{Fermion Mass Hierarchy From Radiative Corrections}},
  \href{https://doi.org/10.1103/PhysRevLett.60.1602}{\emph{Phys. Rev. Lett.}
  {\bfseries 60} (1988) 1602}.

\bibitem{Balakrishna:1988ks}
B.S.~Balakrishna, A.L.~Kagan and R.N.~Mohapatra, \emph{{Quark Mixings and Mass
  Hierarchy From Radiative Corrections}},
  \href{https://doi.org/10.1016/0370-2693(88)91676-0}{\emph{Phys. Lett. B}
  {\bfseries 205} (1988) 345}.

\bibitem{Balakrishna:1988xg}
B.S.~Balakrishna, \emph{{RADIATIVELY INDUCED LEPTON MASSES}},
  \href{https://doi.org/10.1016/0370-2693(88)91480-3}{\emph{Phys. Lett. B}
  {\bfseries 214} (1988) 267}.

\bibitem{Balakrishna:1988bn}
B.S.~Balakrishna and R.N.~Mohapatra, \emph{{Radiative Fermion Masses From New
  Physics at Tev Scale}},
  \href{https://doi.org/10.1016/0370-2693(89)91129-5}{\emph{Phys. Lett. B}
  {\bfseries 216} (1989) 349}.

\bibitem{Babu:1988fn}
K.S.~Babu and X.-G.~He, \emph{{Fermion mass hierarchy and the strong CP
  problem}}, \href{https://doi.org/10.1016/0370-2693(89)90401-2}{\emph{Phys.
  Lett. B} {\bfseries 219} (1989) 342}.

\bibitem{Babu:1989tv}
K.S.~Babu, B.S.~Balakrishna and R.N.~Mohapatra, \emph{{Supersymmetric Model for
  Fermion Mass Hierarchy}},
  \href{https://doi.org/10.1016/0370-2693(90)91433-C}{\emph{Phys. Lett. B}
  {\bfseries 237} (1990) 221}.

\bibitem{Rattazzi:1990wu}
R.~Rattazzi, \emph{{Radiative quark masses constrained by the gauge group
  only}}, \href{https://doi.org/10.1007/BF01562331}{\emph{Z. Phys. C}
  {\bfseries 52} (1991) 575}.

\bibitem{Berezhiani:1991ds}
Z.G.~Berezhiani and R.~Rattazzi, \emph{{Universal seesaw and radiative quark
  mass hierarchy}},
  \href{https://doi.org/10.1016/0370-2693(92)91851-Y}{\emph{Phys. Lett. B}
  {\bfseries 279} (1992) 124}.

\bibitem{Berezhiani:1992bx}
Z.~Berezhiani and R.~Rattazzi, \emph{{Inverted radiative hierarchy of quark
  masses}}, {\emph{JETP Lett.} {\bfseries 56} (1992) 429}.

\bibitem{Berezhiani:1992pj}
Z.G.~Berezhiani and R.~Rattazzi, \emph{{Inverse hierarchy approach to fermion
  masses}}, \href{https://doi.org/10.1016/0550-3213(93)90057-V}{\emph{Nucl.
  Phys. B} {\bfseries 407} (1993) 249}
  [\href{https://arxiv.org/abs/hep-ph/9212245}{{\ttfamily hep-ph/9212245}}].

\bibitem{Arkani-Hamed:1996kxn}
N.~Arkani-Hamed, H.-C.~Cheng and L.J.~Hall, \emph{{A Supersymmetric theory of
  flavor with radiative fermion masses}},
  \href{https://doi.org/10.1103/PhysRevD.54.2242}{\emph{Phys. Rev. D}
  {\bfseries 54} (1996) 2242}
  [\href{https://arxiv.org/abs/hep-ph/9601262}{{\ttfamily hep-ph/9601262}}].

\bibitem{Barr:2007ma}
S.M.~Barr, \emph{{Radiative fermion mass hierarchy in a non-supersymmetric
  unified theory}},
  \href{https://doi.org/10.1103/PhysRevD.76.105024}{\emph{Phys. Rev. D}
  {\bfseries 76} (2007) 105024}
  [\href{https://arxiv.org/abs/0706.1490}{{\ttfamily 0706.1490}}].

\bibitem{Graham:2009gr}
P.W.~Graham and S.~Rajendran, \emph{{A Domino Theory of Flavor}},
  \href{https://doi.org/10.1103/PhysRevD.81.033002}{\emph{Phys. Rev. D}
  {\bfseries 81} (2010) 033002}
  [\href{https://arxiv.org/abs/0906.4657}{{\ttfamily 0906.4657}}].

\bibitem{Dobrescu:2008sz}
B.A.~Dobrescu and P.J.~Fox, \emph{{Quark and lepton masses from top loops}},
  \href{https://doi.org/10.1088/1126-6708/2008/08/100}{\emph{JHEP} {\bfseries
  08} (2008) 100} [\href{https://arxiv.org/abs/0805.0822}{{\ttfamily
  0805.0822}}].

\bibitem{Crivellin:2010ty}
A.~Crivellin, J.~Girrbach and U.~Nierste, \emph{{Yukawa coupling and anomalous
  magnetic moment of the muon: an update for the LHC era}},
  \href{https://doi.org/10.1103/PhysRevD.83.055009}{\emph{Phys. Rev. D}
  {\bfseries 83} (2011) 055009}
  [\href{https://arxiv.org/abs/1010.4485}{{\ttfamily 1010.4485}}].

\bibitem{Crivellin:2011sj}
A.~Crivellin, L.~Hofer, U.~Nierste and D.~Scherer, \emph{{Phenomenological
  consequences of radiative flavor violation in the MSSM}},
  \href{https://doi.org/10.1103/PhysRevD.84.035030}{\emph{Phys. Rev. D}
  {\bfseries 84} (2011) 035030}
  [\href{https://arxiv.org/abs/1105.2818}{{\ttfamily 1105.2818}}].

\bibitem{Adhikari:2015woo}
R.~Adhikari, D.~Borah and E.~Ma, \emph{{New U(1) Gauge Model of Radiative
  Lepton Masses with Sterile Neutrino and Dark Matter}},
  \href{https://doi.org/10.1016/j.physletb.2016.02.039}{\emph{Phys. Lett. B}
  {\bfseries 755} (2016) 414}
  [\href{https://arxiv.org/abs/1512.05491}{{\ttfamily 1512.05491}}].

\bibitem{Chiang:2021pma}
C.-W.~Chiang and K.~Yagyu, \emph{{Radiative Seesaw Mechanism for Charged
  Leptons}}, \href{https://doi.org/10.1103/PhysRevD.103.L111302}{\emph{Phys.
  Rev. D} {\bfseries 103} (2021) L111302}
  [\href{https://arxiv.org/abs/2104.00890}{{\ttfamily 2104.00890}}].

\bibitem{Chiang:2022axu}
C.-W.~Chiang, R.~Obuchi and K.~Yagyu, \emph{{Dark sector as origin of light
  lepton mass and its phenomenology}},
  \href{https://doi.org/10.1007/JHEP05(2022)070}{\emph{JHEP} {\bfseries 05}
  (2022) 070} [\href{https://arxiv.org/abs/2202.07784}{{\ttfamily
  2202.07784}}].

\bibitem{Baker:2020vkh}
M.J.~Baker, P.~Cox and R.R.~Volkas, \emph{{Has the Origin of the Third-Family
  Fermion Masses been Determined?}},
  \href{https://doi.org/10.1007/JHEP04(2021)151}{\emph{JHEP} {\bfseries 04}
  (2021) 151} [\href{https://arxiv.org/abs/2012.10458}{{\ttfamily
  2012.10458}}].

\bibitem{Baker:2021yli}
M.J.~Baker, P.~Cox and R.R.~Volkas, \emph{{Radiative muon mass models and
  $(g-2)_\mu$}}, \href{https://doi.org/10.1007/JHEP05(2021)174}{\emph{JHEP}
  {\bfseries 05} (2021) 174}
  [\href{https://arxiv.org/abs/2103.13401}{{\ttfamily 2103.13401}}].

\bibitem{Yin:2021yqy}
W.~Yin, \emph{{Radiative lepton mass and muon g \ensuremath{-} 2 with
  suppressed lepton flavor and CP violations}},
  \href{https://doi.org/10.1007/JHEP08(2021)043}{\emph{JHEP} {\bfseries 08}
  (2021) 043} [\href{https://arxiv.org/abs/2103.14234}{{\ttfamily
  2103.14234}}].

\bibitem{Chang:2022pue}
W.-F.~Chang, \emph{{Non-universal gauged lepton number for charged lepton
  masses hierarchy and $(g-2)_{e,\mu}$}},
  \href{https://arxiv.org/abs/2210.11097}{{\ttfamily 2210.11097}}.

\bibitem{Zhang:2023zrn}
Y.~Zhang, \emph{{The minimal flavor structure of quarks and leptons}},
  \href{https://doi.org/10.1088/1361-6471/ad074d}{\emph{J. Phys. G} {\bfseries
  50} (2023) 125006} [\href{https://arxiv.org/abs/2302.05943}{{\ttfamily
  2302.05943}}].

\bibitem{Greljo:2023bix}
A.~Greljo and A.E.~Thomsen, \emph{{Rising through the ranks: flavor hierarchies
  from a gauged SU(2) symmetry}},
  \href{https://doi.org/10.1140/epjc/s10052-024-12556-5}{\emph{Eur. Phys. J. C}
  {\bfseries 84} (2024) 213}
  [\href{https://arxiv.org/abs/2309.11547}{{\ttfamily 2309.11547}}].

\bibitem{Arbelaez:2024rbm}
C.~Arbel\'aez, A.E.~C\'arcamo~Hern\'andez, C.~Dib, P.~Escalona~Contreras,
  V.K.~N. and A.~Zerwekh, \emph{{A common framework for fermion mass hierarchy,
  leptogenesis and dark matter}},
  \href{https://arxiv.org/abs/2404.06577}{{\ttfamily 2404.06577}}.

\bibitem{Greljo:2024zrj}
A.~Greljo, A.E.~Thomsen and H.~Tiblom, \emph{{Flavor Hierarchies From SU(2)
  Flavor and Quark-Lepton Unification}},
  \href{https://arxiv.org/abs/2406.02687}{{\ttfamily 2406.02687}}.

\bibitem{Kuchimanchi:2024nkt}
R.~Kuchimanchi, \emph{{Parity and Lepton Masses in the Left Right Symmetric
  Model}},  \href{https://arxiv.org/abs/2406.14480}{{\ttfamily 2406.14480}}.

\bibitem{HernandezGaleana:2004cm}
A.~Hernandez~Galeana and J.H.~Montes~de Oca~Yemha, \emph{{Radiative generation
  of light fermion masses in a SU(3)(H) horizontal symmetry model}},
  {\emph{Rev. Mex. Fis.} {\bfseries 50} (2004) 522}
  [\href{https://arxiv.org/abs/hep-ph/0406315}{{\ttfamily hep-ph/0406315}}].

\bibitem{Reig:2018ocz}
M.~Reig, J.W.F.~Valle and F.~Wilczek, \emph{{SO(3) family symmetry and
  axions}}, \href{https://doi.org/10.1103/PhysRevD.98.095008}{\emph{Phys. Rev.
  D} {\bfseries 98} (2018) 095008}
  [\href{https://arxiv.org/abs/1805.08048}{{\ttfamily 1805.08048}}].

\bibitem{Weinberg:2020zba}
S.~Weinberg, \emph{{Models of Lepton and Quark Masses}},
  \href{https://doi.org/10.1103/PhysRevD.101.035020}{\emph{Phys. Rev. D}
  {\bfseries 101} (2020) 035020}
  [\href{https://arxiv.org/abs/2001.06582}{{\ttfamily 2001.06582}}].

\bibitem{Jana:2021tlx}
S.~Jana, S.~Klett and M.~Lindner, \emph{{Flavor seesaw mechanism}},
  \href{https://doi.org/10.1103/PhysRevD.105.115015}{\emph{Phys. Rev. D}
  {\bfseries 105} (2022) 115015}
  [\href{https://arxiv.org/abs/2112.09155}{{\ttfamily 2112.09155}}].

\bibitem{Mohanta:2022seo}
G.~Mohanta and K.M.~Patel, \emph{{Radiatively generated fermion mass hierarchy
  from flavor nonuniversal gauge symmetries}},
  \href{https://doi.org/10.1103/PhysRevD.106.075020}{\emph{Phys. Rev. D}
  {\bfseries 106} (2022) 075020}
  [\href{https://arxiv.org/abs/2207.10407}{{\ttfamily 2207.10407}}].

\bibitem{Mohanta:2023soi}
G.~Mohanta and K.M.~Patel, \emph{{Gauged SU(3)$_{F}$ and loop induced quark and
  lepton masses}}, \href{https://doi.org/10.1007/JHEP10(2023)128}{\emph{JHEP}
  {\bfseries 10} (2023) 128}
  [\href{https://arxiv.org/abs/2308.05642}{{\ttfamily 2308.05642}}].

\bibitem{Berezhiani:1983hm}
Z.G.~Berezhiani, \emph{{The Weak Mixing Angles in Gauge Models with Horizontal
  Symmetry: A New Approach to Quark and Lepton Masses}},
  \href{https://doi.org/10.1016/0370-2693(83)90737-2}{\emph{Phys. Lett. B}
  {\bfseries 129} (1983) 99}.

\bibitem{Berezhiani:1985in}
Z.G.~Berezhiani, \emph{{Horizontal Symmetry and Quark - Lepton Mass Spectrum:
  The SU(5) x SU(3)-h Model}},
  \href{https://doi.org/10.1016/0370-2693(85)90164-9}{\emph{Phys. Lett. B}
  {\bfseries 150} (1985) 177}.

\bibitem{Chang:1986bp}
D.~Chang and R.N.~Mohapatra, \emph{{Small and Calculable Dirac Neutrino Mass}},
  \href{https://doi.org/10.1103/PhysRevLett.58.1600}{\emph{Phys. Rev. Lett.}
  {\bfseries 58} (1987) 1600}.

\bibitem{FernandezNavarro:2023rhv}
M.~Fern\'andez~Navarro and S.F.~King, \emph{{Tri-hypercharge: a separate gauged
  weak hypercharge for each fermion family as the origin of flavour}},
  \href{https://doi.org/10.1007/JHEP08(2023)020}{\emph{JHEP} {\bfseries 08}
  (2023) 020} [\href{https://arxiv.org/abs/2305.07690}{{\ttfamily
  2305.07690}}].

\bibitem{Davighi:2023evx}
J.~Davighi and B.A.~Stefanek, \emph{{Deconstructed hypercharge: a natural model
  of flavour}}, \href{https://doi.org/10.1007/JHEP11(2023)100}{\emph{JHEP}
  {\bfseries 11} (2023) 100}
  [\href{https://arxiv.org/abs/2305.16280}{{\ttfamily 2305.16280}}].

\bibitem{UTfit:2007eik}
{\scshape UTfit} collaboration, \emph{{Model-independent constraints on $\Delta
  F=2$ operators and the scale of new physics}},
  \href{https://doi.org/10.1088/1126-6708/2008/03/049}{\emph{JHEP} {\bfseries
  03} (2008) 049} [\href{https://arxiv.org/abs/0707.0636}{{\ttfamily
  0707.0636}}].

\bibitem{Ciuchini:1998ix}
M.~Ciuchini et~al., \emph{{Delta M(K) and epsilon(K) in SUSY at the
  next-to-leading order}},
  \href{https://doi.org/10.1088/1126-6708/1998/10/008}{\emph{JHEP} {\bfseries
  10} (1998) 008} [\href{https://arxiv.org/abs/hep-ph/9808328}{{\ttfamily
  hep-ph/9808328}}].

\bibitem{Becirevic:2001jj}
D.~Becirevic, M.~Ciuchini, E.~Franco, V.~Gimenez, G.~Martinelli, A.~Masiero
  et~al., \emph{{$B_d - \bar{B}_d$ mixing and the $B_d \to J/\psi K_s$
  asymmetry in general SUSY models}},
  \href{https://doi.org/10.1016/S0550-3213(02)00291-2}{\emph{Nucl. Phys. B}
  {\bfseries 634} (2002) 105}
  [\href{https://arxiv.org/abs/hep-ph/0112303}{{\ttfamily hep-ph/0112303}}].

\bibitem{Kitano:2002mt}
R.~Kitano, M.~Koike and Y.~Okada, \emph{{Detailed calculation of lepton flavor
  violating muon electron conversion rate for various nuclei}},
  \href{https://doi.org/10.1103/PhysRevD.76.059902}{\emph{Phys. Rev. D}
  {\bfseries 66} (2002) 096002}
  [\href{https://arxiv.org/abs/hep-ph/0203110}{{\ttfamily hep-ph/0203110}}].

\bibitem{Smolkovic:2019jow}
A.~Smolkovi\v{c}, M.~Tammaro and J.~Zupan, \emph{{Anomaly free Froggatt-Nielsen
  models of flavor}},
  \href{https://doi.org/10.1007/JHEP10(2019)188}{\emph{JHEP} {\bfseries 10}
  (2019) 188} [\href{https://arxiv.org/abs/1907.10063}{{\ttfamily
  1907.10063}}].

\bibitem{SINDRUMII:2006dvw}
{\scshape SINDRUM II} collaboration, \emph{{A Search for muon to electron
  conversion in muonic gold}},
  \href{https://doi.org/10.1140/epjc/s2006-02582-x}{\emph{Eur. Phys. J. C}
  {\bfseries 47} (2006) 337}.

\bibitem{Heeck:2016xkh}
J.~Heeck, \emph{{Lepton flavor violation with light vector bosons}},
  \href{https://doi.org/10.1016/j.physletb.2016.05.007}{\emph{Phys. Lett. B}
  {\bfseries 758} (2016) 101}
  [\href{https://arxiv.org/abs/1602.03810}{{\ttfamily 1602.03810}}].

\bibitem{Lavoura:2003xp}
L.~Lavoura, \emph{{General formulae for f(1) ---\ensuremath{>} f(2) gamma}},
  \href{https://doi.org/10.1140/epjc/s2003-01212-7}{\emph{Eur. Phys. J. C}
  {\bfseries 29} (2003) 191}
  [\href{https://arxiv.org/abs/hep-ph/0302221}{{\ttfamily hep-ph/0302221}}].

\bibitem{Capozzi:2017ipn}
F.~Capozzi, E.~Di~Valentino, E.~Lisi, A.~Marrone, A.~Melchiorri and A.~Palazzo,
  \emph{{Global constraints on absolute neutrino masses and their ordering}},
  \href{https://doi.org/10.1103/PhysRevD.95.096014}{\emph{Phys. Rev. D}
  {\bfseries 95} (2017) 096014}
  [\href{https://arxiv.org/abs/2003.08511}{{\ttfamily 2003.08511}}].

\bibitem{deSalas:2020pgw}
P.F.~de~Salas, D.V.~Forero, S.~Gariazzo, P.~Mart\'\i{}nez-Mirav\'e, O.~Mena,
  C.A.~Ternes et~al., \emph{{2020 global reassessment of the neutrino
  oscillation picture}},
  \href{https://doi.org/10.1007/JHEP02(2021)071}{\emph{JHEP} {\bfseries 02}
  (2021) 071} [\href{https://arxiv.org/abs/2006.11237}{{\ttfamily
  2006.11237}}].

\bibitem{Esteban:2020cvm}
I.~Esteban, M.C.~Gonzalez-Garcia, M.~Maltoni, T.~Schwetz and A.~Zhou,
  \emph{{The fate of hints: updated global analysis of three-flavor neutrino
  oscillations}}, \href{https://doi.org/10.1007/JHEP09(2020)178}{\emph{JHEP}
  {\bfseries 09} (2020) 178}
  [\href{https://arxiv.org/abs/2007.14792}{{\ttfamily 2007.14792}}].

\bibitem{Weinberg:1979sa}
S.~Weinberg, \emph{{Baryon and Lepton Nonconserving Processes}},
  \href{https://doi.org/10.1103/PhysRevLett.43.1566}{\emph{Phys. Rev. Lett.}
  {\bfseries 43} (1979) 1566}.

\bibitem{Minkowski:1977sc}
P.~Minkowski, \emph{{$\mu \to e\gamma$ at a Rate of One Out of $10^{9}$ Muon
  Decays?}}, \href{https://doi.org/10.1016/0370-2693(77)90435-X}{\emph{Phys.
  Lett. B} {\bfseries 67} (1977) 421}.

\bibitem{Yanagida:1979as}
T.~Yanagida, \emph{{Horizontal gauge symmetry and masses of neutrinos}},
  {\emph{Conf. Proc. C} {\bfseries 7902131} (1979) 95}.

\bibitem{Mohapatra:1979ia}
R.N.~Mohapatra and G.~Senjanovic, \emph{{Neutrino Mass and Spontaneous Parity
  Nonconservation}},
  \href{https://doi.org/10.1103/PhysRevLett.44.912}{\emph{Phys. Rev. Lett.}
  {\bfseries 44} (1980) 912}.

\bibitem{Schechter:1980gr}
J.~Schechter and J.W.F.~Valle, \emph{{Neutrino Masses in SU(2) x U(1)
  Theories}}, \href{https://doi.org/10.1103/PhysRevD.22.2227}{\emph{Phys. Rev.
  D} {\bfseries 22} (1980) 2227}.

\bibitem{Foot:1988aq}
R.~Foot, H.~Lew, X.G.~He and G.C.~Joshi, \emph{{Seesaw Neutrino Masses Induced
  by a Triplet of Leptons}}, \href{https://doi.org/10.1007/BF01415558}{\emph{Z.
  Phys. C} {\bfseries 44} (1989) 441}.

\bibitem{Lazarides:1980nt}
G.~Lazarides, Q.~Shafi and C.~Wetterich, \emph{{Proton Lifetime and Fermion
  Masses in an SO(10) Model}},
  \href{https://doi.org/10.1016/0550-3213(81)90354-0}{\emph{Nucl. Phys. B}
  {\bfseries 181} (1981) 287}.

\bibitem{Magg:1980ut}
M.~Magg and C.~Wetterich, \emph{{Neutrino Mass Problem and Gauge Hierarchy}},
  \href{https://doi.org/10.1016/0370-2693(80)90825-4}{\emph{Phys. Lett. B}
  {\bfseries 94} (1980) 61}.

\bibitem{Mohapatra:1980yp}
R.N.~Mohapatra and G.~Senjanovic, \emph{{Neutrino Masses and Mixings in Gauge
  Models with Spontaneous Parity Violation}},
  \href{https://doi.org/10.1103/PhysRevD.23.165}{\emph{Phys. Rev. D} {\bfseries
  23} (1981) 165}.

\bibitem{Georgi:1974sy}
H.~Georgi and S.L.~Glashow, \emph{{Unity of All Elementary Particle Forces}},
  \href{https://doi.org/10.1103/PhysRevLett.32.438}{\emph{Phys. Rev. Lett.}
  {\bfseries 32} (1974) 438}.

\bibitem{Fritzsch:1974nn}
H.~Fritzsch and P.~Minkowski, \emph{{Unified Interactions of Leptons and
  Hadrons}}, \href{https://doi.org/10.1016/0003-4916(75)90211-0}{\emph{Annals
  Phys.} {\bfseries 93} (1975) 193}.

\bibitem{Pati:1974yy}
J.C.~Pati and A.~Salam, \emph{{Lepton Number as the Fourth Color}},
  \href{https://doi.org/10.1103/PhysRevD.10.275}{\emph{Phys. Rev. D} {\bfseries
  10} (1974) 275}.

\bibitem{Holdom:1985ag}
B.~Holdom, \emph{{Two U(1)'s and Epsilon Charge Shifts}},
  \href{https://doi.org/10.1016/0370-2693(86)91377-8}{\emph{Phys. Lett. B}
  {\bfseries 166} (1986) 196}.

\bibitem{ATLAS:2015yey}
{\scshape ATLAS, CMS} collaboration, \emph{{Combined Measurement of the Higgs
  Boson Mass in $pp$ Collisions at $\sqrt{s}=7$ and 8 TeV with the ATLAS and
  CMS Experiments}},
  \href{https://doi.org/10.1103/PhysRevLett.114.191803}{\emph{Phys. Rev. Lett.}
  {\bfseries 114} (2015) 191803}
  [\href{https://arxiv.org/abs/1503.07589}{{\ttfamily 1503.07589}}].

\bibitem{Xing:2007fb}
Z.-z.~Xing, H.~Zhang and S.~Zhou, \emph{{Updated Values of Running Quark and
  Lepton Masses}},
  \href{https://doi.org/10.1103/PhysRevD.77.113016}{\emph{Phys. Rev. D}
  {\bfseries 77} (2008) 113016}
  [\href{https://arxiv.org/abs/0712.1419}{{\ttfamily 0712.1419}}].

\bibitem{ParticleDataGroup:2022pth}
{\scshape Particle Data Group} collaboration, \emph{{Review of Particle
  Physics}}, \href{https://doi.org/10.1093/ptep/ptac097}{\emph{PTEP} {\bfseries
  2022} (2022) 083C01}.

\bibitem{Mummidi:2021anm}
V.S.~Mummidi and K.M.~Patel, \emph{{Leptogenesis and fermion mass fit in a
  renormalizable SO(10) model}},
  \href{https://doi.org/10.1007/JHEP12(2021)042}{\emph{JHEP} {\bfseries 12}
  (2021) 042} [\href{https://arxiv.org/abs/2109.04050}{{\ttfamily
  2109.04050}}].

\end{thebibliography}\endgroup
\bibliographystyle{JHEP.bst}
\end{document}